%% file: GKAT.tex
\documentclass[11pt,twoside]{article}

\usepackage{fancyhdr}
\usepackage{epsfig}
\usepackage{latexsym}
\usepackage{microtype}
\usepackage{lastpage} 
\usepackage{hyperref}
\usepackage{graphicx}
\usepackage{lipsum}

\usepackage[dvipsnames*,svgnames]{xcolor}
\hypersetup{colorlinks=true,allcolors=DarkBlue,linkcolor=black}

\usepackage[utf8]{inputenc}
\usepackage{amssymb}
\usepackage{thmtools,amsmath}
\usepackage[all,cmtip]{xy}
\usepackage{arydshln}

\usepackage{color}
\usepackage{pdfpages}
\usepackage{textcomp}

\usepackage{amscd}
\usepackage{stmaryrd}

\usepackage{fancyvrb}
\usepackage{boxedminipage}
\usepackage{listings}
\usepackage{minitoc}
\input xy
\xyoption{all}
\input{dinmacros.tex}

\usepackage{mathtools}
\usepackage{varwidth}
\usepackage{float}

\newcommand\myeq{\mathrel{\overset{\makebox[0pt]{\mbox{\normalfont\tiny\sffamily def}}}{=}}}

\newtheorem{theorem}{Theorem}

\newtheorem{definition}{Definition}
\newtheorem{lemma}{Lemma}

\newtheorem{example}{Example}

\newcommand{\proof}[1]{\par\noindent {\bf Proof:}\ \ \ #1 \hfill$\Box$\hspace{1ex}}

\allowdisplaybreaks[3]

\allowdisplaybreaks[3]

\usepackage[normalem]{ulem}

\newcommand{\TheAuthor}{}
\newcommand{\Author}[1]{\renewcommand{\TheAuthor}{#1}}

\newcommand{\TheTitle}{}
\newcommand{\Title}[1]{\renewcommand{\TheTitle}{#1}}

\Author{Gomes, Madeira, Barbosa}
\Title{Generalising KAT to verify weighted computations}
\begin{document}


\parindent=8mm
\noindent

\vskip -3mm
\noindent
\vskip -1mm
\noindent

\vspace{1cm}
\begin{center}
{\Large\bf Generalising KAT to verify weighted computations
}
\end{center}
\vspace{4mm}

\begin{center}
  {\large Leandro GOMES}\footnote{HASLab INESC TEC, Universidade do Minho,
   R. da Universidade, 4710-057 Braga, Portugal, Email: {\tt leandro.r.gomes@inesctec.pt}},
   {\large Alexandre MADEIRA}\footnote{CIDMA, Universidade de Aveiro,
   Campus Universitário de Santiago, 3810-193 Aveiro, Portugal, Email: {\tt madeira@ua.pt}},
   {\large Luis S. BARBOSA}\footnote{HASLab INESC TEC, Universidade do Minho,
   R. da Universidade, 4710-057 Braga, Portugal \& QuantaLab, INL, Braga, Portugal,  Email: {\tt lsb@di.uminho.pt}} 
  \end{center}
  \vspace{3ex}
  
  \date{}

  \begin{abstract}
    Kleene algebra with tests (KAT) was introduced as an algebraic structure to model and reason about classic imperative programs, i.e. sequences of discrete transitions guarded by Boolean tests. This paper introduces two generalisations of this structure able to express programs as weighted transitions and tests with outcomes in non necessarily bivalent truth spaces: 
    graded Kleene algebra with tests (GKAT) and a variant where tests are also idempotent (I-GKAT). 
    On this context, and in analogy to Kozen's encoding of Propositional Hoare Logic (PHL) in 
    KAT \cite{Kozen2000}, we discuss the encoding of a graded PHL in I-GKAT and of its while-free fragment in GKAT.
    Moreover, to establish semantics for these structures four new algebras are defined: $\boldsymbol{FSET(T)}$, $\boldsymbol{FREL(K,T)}$ and $\boldsymbol{FLANG(K,T)}$ over complete residuated lattices $\boldsymbol{K}$ and $\boldsymbol{T}$, and $\boldsymbol{M(n,A)}$ over a GKAT or I-GKAT $\boldsymbol{\A}$. 
    As a final exercise, the paper discusses some program equivalence proofs in a graded context.
    
    \smallskip

\noindent
{\bf Keywords:} $Kleene\ algebra$, $Hoare\ logic$, $graded\ tests$, $fuzzy\ relations$

    \end{abstract}



    
 


\section{Introduction}\label{sec:intro} 

\subsection{Roadmap}\label{subsec:roadmap}

\input{intro}

\newpage
\section{Graded Kleene Algebra with Tests}\label{sec:graded_kleene_algebra_with_tests}

\subsection{The basic structure}\label{subsec:basic_structure}

\input{prelim}

\subsection{Graded Propositional Hoare Logic}
\label{subsec:hoare_logic}
\input{hoare}
\newpage
\section{Idempotent graded Kleene Algebra with Tests}
\label{sec:heyting}
\subsection{The basic structure}\label{subsec:basic_structure2}
\input{heyting}

\newpage
\section{Illustration: Fuzzy Sets, Fuzzy Relations, Fuzzy Languages and Matrices as GKAT/I-GKAT}
\label{sec:fuzzy}
\input{fuzzy}
\input{matrices}

\newpage
\section{A folk theorem adapted to a graded scenario}
\label{sec:folk}
\input{folk}

%
%
%
 
\subsection{Nested Loops}\label{subsec:nloops}
\input{loops}
 
\newpage
\section{Conclusion and Further Work} 
\label{sec:conclusion_and_further_work}
\input{conclusions}
\newpage
\section*{Acknowledgements}
\label{thanks}
 This work is financed by the ERDF – European Regional Development Fund through the Operational Programme for Competitiveness and Internationalisation - COMPETE 2020 Programme and by National Funds through the Portuguese funding agency, FCT - Funda\c{c}\~ao para a Ci\^encia e a Tecnologia, within projects \texttt{POCI-01-0145-FEDER-030947} and \texttt{UID/MAT/04106/2019}. The second author is supported in the scope of the framework contract foreseen in the numbers 4, 5 and 6 of the article 23, of the Decree-Law 57/2016, of August 29, changed by Portuguese Law 57/2017, of July 19. This paper is also a result of the project SmartEGOV: Harnessing EGOV for Smart Governance (Foundations, Methods, Tools) NORTE-01-0145-FEDER-000037, supported by Norte Portugal Regional Operational Programme (NORTE 2020), under the PORTUGAL 2020 Partnership Agreement, through the European Regional Development Fund (EFDR). It received further support from the PT-FLAD Chair in Smart Cities \& Smart Governance.


\newpage

\bibliographystyle{plainurl}
\bibliography{thesis.bib}

\newpage
\appendix
\renewcommand{\thesection}{\Alph{section}}
\input{appendices}


\end{document}

%% file: dinmacros.tex
\usepackage{amsmath}

\usepackage{mathtools}

\def\arrayin#1{\begin{array}{rcl}#1\end{array}}

\def\A{\mathbf{A}}

\newcommand{\T}{\mathbf{T}}
\newcommand{\K}{\mathbf{K}}

\newcommand{\REL}{\REL}

\def\square{\Box}

\def\PL{\mathit{PL}}
\def\REL{\mathit{REL}}

\def\H2PL{\mathcal{H}^2\PL}

\def\rcb#1#2#3#4{\def\nothing{}\def\range{#3}\mathopen{\langle}#1 \ #2 \ \ifx\range\nothing::\else: \ #3 :\fi \ #4\mathclose{\rangle}}

 \def\just#1#2{\\ &#1& \rule{2em}{0pt} \{
\mbox{\rule[-.7em]{0pt}{1.8em} \footnotesize{#2}} \} \\ && }

\DeclareMathAlphabet{\mathbb}{U}{msb}{m}{n}
\DeclareSymbolFont{ams}{U}{msa}{m}{n}
\DeclareSymbolFontAlphabet{\mathams}{ams}
\DeclareMathSymbol{\filter}{\mathams}{ams}{22}

%% file: intro.tex

Kleene algebra is pervasive in Computer Science, applications ranging from semantics and logics of programs, to automata and formal language theory, as well as to the design and analysis of algorithms. Some recent examples deal with hybrid systems analysis \cite{HofnerM09}, separation logic \cite{DANG2011221} and non-termination analysis \cite{termination}.
As a program calculus, the axiomatisation of Kleene algebra forms a deductive system to manipulate programs \cite{Kozen1994}.
Its applications typically deal with conventional, imperative programming constructs, namely conditionals and loops. Reasoning equationally about them entails the need for a notion of a test, which leads to the development of \emph{Kleene algebra with tests} (KAT) \cite{Kozen1997} combining the expressiveness of Kleene algebra with a Boolean subalgebra to formalise tests. An alternative approach extends a Kleene algebra with both a domain and a codomain operation mapping transitions to propositions \cite{DesharnaisMS06}. Contrary to KAT, the resulting structure is a one-sorted algebra.
D. Kozen \cite{Kozen1994} proved that plain Kleene algebra is closed under the formation of square matrices, later 
extending this result to Kleene algebra with tests by considering a test a Boolean diagonal matrix.

\emph{Hoare logic} (HL) was the first formal system proposed for verification of programs. Introduced in 1969, its wide influence made it a cornerstone in program correctness. HL encompasses a syntax to reason about partial correctness assertions of the form $\{b\}p\{c\}$, called a Hoare triple, and a deductive system to reason about them \cite{Hoare69}, \cite{Floyd1993}. In a Hoare triple, $b$ and $c$ stand for predicates, representing the \emph{pre} and \emph{post} conditions, respectively, and $p$ is a program statement.
\emph {Propositional Hoare logic} (PHL) is a fragment of HL, in which Hoare triples are reduced to static assertions about the underlying domain of computation \cite{Kozen2000}, and therefore encoded in a Kleene algebra with tests. The translation maps Hoare triples to equations and the rules of inference into equational implications.

As originally presented, KAT is suitable to reason about classic imperative programs. In fact, such programs are particularly ``well tractable'': they represent a sequence of discrete steps, each of one can be modelled as an atomic transition in a standard automaton. Typically, these assertions have an outcome in a bivalent truth space.
However, current complex, dynamic systems require new computing domains, namely probabilistic \cite{Qiao2008} or continuous \cite{Platzer}, which entail the need for computing paradigms able to deal with some sort of weighted program executions. Actually, assertions about these programs have often a graded outcome.

In this context, the development of algebraic structures to model weighted computations becomes a must. Such computations, often associated with notions of uncertainty, can be mathematically conceptualised in terms of the well established fuzzy set theory.
Although a fuzzy set was initially defined as a mapping from a set $X$ to the unit interval $[0,1]$ \cite{ZADEH1965338}, it later evolved into a more generic concept, by replacing such an interval by an arbitrary complete distributive lattice $L$ \cite{GOGUEN1967145}.
The later work constitutes a cornerstone in the study of algebraic formalisations of fuzzy concepts. M. Winter \cite{Winter00,Winter01,Winter11} follows this route through a categorical perspective. The work of J. Desharnais \emph{et al.} \cite{DesharnaisMS06} continued along distinct paths: one \cite{DesharnaisS11} proposes a new axiomatisation for domain and codomain operators, leading to algebras of domain elements of which Boolean and Heyting algebras are special cases; another \cite{DesharnaisM14} investigates notions of domain and codomain operators to provide applications in fuzzy relations and matrices, by using an idempotent left semiring as the base algebraic structure.

This paper builds on such motivations to introduce two generalisations of KAT able to express programs as weighted computations and tests as predicates evaluated in a graded truth space - the \emph{graded Kleene algebra with tests} (GKAT) and the \emph{idempotent graded Kleene algebra with tests} (I-GKAT).
GKAT has several interesting instances, from the continuous \L ukasiewicz lattice to the discrete finite hoops. I-GKAT, on the other hand, is able to encode, with the exception of the assignment rule, the deductive system of PHL.
In analogy to KAT \cite{Kozen2000}, we discuss how to encode PHL into GKAT, therefore extending the classical scope of program correctness.
However, this can only be entirely achieved for the fragment of $while$-free programs.
To obtain a complete encoding of Hoare logic, there was a need to refine the basic structure. Thus, I-GKAT emerged as
 a subclass of GKAT, with, of course a smaller set of instances.
 This includes, in particular, lattice $\mathbf{3}$ to deal with partial programs and uncertainty on tests, and G\"odel algebra, a well-known structure used in logics whose truth values are closed subsets of the interval $[0,1]$.

Extending KAT to the domain of weighted computation is the main motivation of this work. The paper extends some preliminary results documented in our previous work \cite{GKAT} in distinct directions. First, we propose three algebraic constructions that represent models for both GKAT and I-GKAT: the set of all fuzzy sets, the set of all fuzzy relations and the set of all fuzzy languages, provided with the appropriate operators over the elements for each case. Note that
in modelling uncertainty fuzzy logic plays a very important role. It is known that the standard algebraic model for classic, bivalent logic, is a Boolean algebra, with a clear connection to the classic set theory. Similarly, as stated in \cite{MPFL}, reasoning with uncertainty, as captured by fuzzy logic, is tied to fuzzy set theory.

At a latter stage, we prove that both GKAT and I-GKAT enjoy a matricial construction similar to D. Kozen's classical result \cite{Kozen1994}. This is indeed relevant as many problems modelled as labelled transition systems can be formulated as matrices over a Klenee algebra or a similar structure. Constructions are parametric on the concrete underlying lattice, as defined by R. Guillherme \cite{Guilherme2016} for the case of fuzzy sets, relations and languages.
Finally, we revisit in the weighted context some examples of equational proofs from the KAT seminal paper \cite{Kozen1997}. In particular, we show how to handle, in such a scenario, the result of denesting two nested \textbf{while} loops.



 The remainder of the paper is organised as follows: Subsection \ref{sub:kleene_algebra_with_tests_vs_hoare_logic} recapitulates some fundamental concepts. Section \ref{sec:graded_kleene_algebra_with_tests} introduces \emph{graded Kleene algebra with tests} as a generalisation of KAT, detailing its axiomatisation, a few examples and proofs of basic properties. It also presents a partial encoding of classical PHL in GKAT.
 Section \ref{sec:heyting} introduces \emph{idempotent graded Kleene algebra with tests} as another generalisation of the standard KAT and a refinement of GKAT, offering a complete encoding of PHL. Section \ref{sec:fuzzy} presents the sets of all fuzzy sets, fuzzy relations, fuzzy languages and $n\times n$ matrices, with the appropriate operations, as models of GKAT and I-GKAT. 
In Section \ref{sec:folk} we discuss some equational proofs for program equivalences in a graded scenario.
Finally, Section \ref{sec:conclusion_and_further_work} sums up some related research, concludes, and enumerates some topics for future work.


\subsection{Preliminaries} 
\label{sub:kleene_algebra_with_tests_vs_hoare_logic}
\begin{definition}\label{def:KleeneAlgebra}
	A \emph{Kleene algebra with tests} (KAT) is a tuple
	\[(K,T,+,;,^*,\bar{}\ ,0,1)\]
	where $T\subseteq K$, $0$ and $1$ are constants in $T$, $+$ and $;$ are binary operators in both $K$ and $T$, $^*$ is a unary operator in $K$, and $\ \bar{}\ $ is a unary operator defined only on $T$ such that:	
	\begin{itemize}
	\item $(K,+,;,^*,0,1)$ is a Kleene algebra;
	\item $(T,+,;,\bar{}\ ,0,1)$ is a Boolean algebra;
	\item $(T,+,;,0,1)$ is a subalgebra of $(K,+,;,0,1)$.
	\end{itemize}
	
\end{definition}
The elements of $K$, denoted by lower case letters $p,q,r,s,x,y,z$, stand for programs and the elements of $T$, denoted by $a,b,c,d$ are called tests.
Kleene algebra with tests induces an abstract programming language, where conditionals and \textbf{while} loops programming constructs are encoded as follows:
\begin{align*}
\textbf{if}\ b\ \textbf{then}\ p\ \myeq b;p+\bar{b}\\
\textbf{if}\ b\ \textbf{then}\ p\ \textbf{else}\ q \myeq b;p+\bar{b};q\\
\textbf{while}\ b\ \textbf{do}\ p \myeq (b;p)^*;\bar{b}
\end{align*}
The encoding of Propositional Hoare Logic (PHL) in KAT leads to an equational calculus to reason about Hoare triples. Recall that one such triple $\{b\}p\{c\}$ is valid if whenever precondition $b$ is met, the postcondition $c$ is guaranteed to hold, upon the successful termination of program $p$. Classically, validity in PHL is established through the set of rules in Figure \ref{fig:rules}.
		\begin{figure}[ht]
\centering
\begin{minipage}{\linewidth}



\begin{itemize}

\item \textit{Composition rule}:

\begin{equation*}
\frac{\{b\}p\{c\}\;\;\;\{c\}q\{d\}}{\{b\}p;q\{d\}}\label{eq:comprule}
\end{equation*}

\item \textit{Conditional rule}:

\begin{equation*}
\frac{\{b\wedge c\}p\{d\}, \{\neg b\wedge c\}q\{d\}}{\{c\}\ \textbf{if} \ b\ \textbf{then}\ p\ \textbf{else}\ q\ \{d\}}\label{eq:condrule}
\end{equation*}

\end{itemize}

	
\begin{itemize}

\item \textit{While rule}:

\begin{equation*}
\frac{\{b\wedge c\}p\{c\}}{\{c\}\ \textbf{while} \ b\ \textbf{do}\ p\{\neg b \wedge c\}}\label{eq:whilerule}
\end{equation*}

\item \textit{Weakening and Strengthening rule}:

\begin{equation*}
\frac{b'\rightarrow b,\ \{b\}p\{c\},\ c\rightarrow c'}{\{b'\}\ p \{c'\}}\label{eq:weakrule}
\end{equation*}

\end{itemize}


\end{minipage}

\caption{Hoare logic rules.}
\label{fig:rules}
\end{figure}

A Hoare triple $\{b\}p\{c\}$ is encoded in KAT as $b;p;\bar{c}= 0 \label{eq:pca}$, which is equivalent to $b;p=b;p;c \label{eq:pca2}$.
The first equation means, intuitively, that the execution of $p$ with precondition $b$ and postcondition $\bar{c}$ does not halt.
Equation $b;p=b;p;c$, on the other hand, states that the verification of the post condition $c$ after the execution of $b;p$ is redundant.
PHL inference rules are encoded in KAT, as follows:

\begin{itemize}
\item \textit{Composition}:
\begin{equation*}
b;p= b;p;c \wedge c;q= c;q;d \Rightarrow b;p;q= b;p;q;d\label{eq:comprulekat}
\end{equation*}
\item \textit{Conditional}:
\begin{equation*}
b;c;p= b;c;p;d \wedge \bar{b};c;q= \bar{b};c;q;d
\Rightarrow c;(b;p + \bar{b};q) = c;(b;p + \bar{b};q);d\label{eq:condrulekat}
\end{equation*}
\item \textit{While}:
\begin{equation*}
b;c;p = b;c;p;c \Rightarrow c;(b;p)^*; \bar{b} = c;(b;p)^*;\bar{b}; \bar{b};c\label{eq:whilerulekat}
\end{equation*}
\item \textit{Weakening and Strengthening}:
\begin{equation*}
b'\leq b \wedge b;p= b;p;c \wedge c\leq c' \Rightarrow b';p= b';p;c'\label{eq:weakrulekat}
\end{equation*}
\end{itemize}

where $\leq$ refers to the partial order on $K$ defined as $p\leq q$ iff $p+q=q$.


%% file: prelim.tex




The approach proposed in this paper, to reason about program executions in a weighted, i.e. many-valued context, is based on redefining the interpretation of the assertions about programs. Since such assertions take the form of tests, we start by modifying the part of the axiomatisation of KAT that deals with properties of tests, i.e. the Boolean algebra $(T,+,\cdot,\bar{}\ ,0,1)$.

Instead of having a Boolean outcome, as in KAT, tests are graded, taking values from a truth space with more than two possible outcomes. As a consequence, the expression $b;p$ represents a weighted execution of program $p$, guarded by the value of test $b$. This leads to the following generalisation of KAT:
 \begin{definition}\label{def:gradedKleeneAlgebra}
 A \emph{graded Kleene algebra with tests} (GKAT) is a tuple
	\[(K,T,+,;,^*,\rightarrow,0,1)\]
	where $K$ and $T$ are sets, with $T\subseteq K$, $0$ and $1$ are constants in $T$, $+$ and $;$ are binary operations in both $K$ and $T$, $^*$ is a unary operator in $K$, and $\rightarrow$ is an operator only defined in $T$, satisfying the axioms in Figure \ref{fig:ax}. Relation $\leq$ is induced by $+$ in the usual way: $p\leq q$ iff $p+q=q$.
	
 \end{definition}
Again, programs are elements of $K$ denoted by lower case letters $p,q,r,s,x,y,z$ and tests are elements of $T$ denoted by $a,b,c,d$.
Observe that a Kleene algebra is recovered by restricting the definition of \emph{GKAT} to $(K,T,+,;,^*,0,1)$, axiomatised by (\ref{eqn1})-(\ref{eqn13}).
Note also that $(T,+,;,0,1)$ is a subalgebra of $(K,+,;,0,1)$. Differently from what happens in KAT, negation $\bar{a}$, for $a\in T$, is not explicitly denoted, although it can be derived as $a\rightarrow 0$.


 
		\begin{figure}[ht]
\begin{minipage}{0.5\linewidth}
			\begin{eqnarray}
				p+ (q +r) & = & (p+ q) +r \label{eqn1}\\
				p +q & = & q + p \label{eqn2}\\
				p;(q;r) & = & (p;q);r \label{eqn5}\\
				p;1 & = & 1;p = p \label{eqn6}\\
				p;(q +r ) & =& (p;q) + (p;r) \label{eqn7}\\
				(p +q);r & =& (p;r) + (q;r) \;\;\label{eqn8}\\
				p;0 & = & 0;p = 0 \label{eqn9}
				\end{eqnarray}
\end{minipage}
\quad
\begin{minipage}{0.50\linewidth}
\begin{eqnarray}
1+p;p^* & = & p^* \label{eqn10}\\
q+p;r \leq r & \Rightarrow & p^*;q \leq r \label{eqn12}\\
				q+r;p \leq r & \Rightarrow & q;p^* \leq r \label{eqn13}\\
				a;b \leq c & \Leftrightarrow & b \leq a \rightarrow c \label{eqn14} \\
				a &\leq & 1 \label{eqn17}\\
				a;b & = & b;a \label{eqn18}
			\end{eqnarray}
\end{minipage}
\caption{Axiomatisation of graded Kleene algebra with tests.}
\label{fig:ax}
\end{figure}

Note that a Kleene algebra is usually characterised by three more equations:

\begin{eqnarray}
p +p & = & p \label{eqn3}\\
p + 0 & = & p\label{eqn4}\\
1+p^*;p & = & p^*\label{eqn11}
\end{eqnarray}
We resort to these equations to prove some results of this paper. However, as can be easily veryfied, they can be derived from the axiomatisation of Figure \ref{fig:ax}.

Operators ``$+$'' and ``$;$'' in GKAT play a different role when acting on programs or tests. The former stands for non-deterministic choice over programs, and a form of logical disjunction on tests. The latter is taken as sequential composition of actions when applied to elements of $K$, and as a "multiplication" of tests when applied to elements of $T$. Finally, in the domain of programs, the constants $0$ and $1$ interpret the \emph{halt} and \emph{skip} commands, while when applied to tests, stand for logical constants false and true, respectively. Some operations are specific to only tests or programs. For instance, operation $^*$ stands for iterative execution of programs and operation $\rightarrow$ plays the role of logical implication over tests.



A main particularity of the GKAT axiomatization concerns rules (\ref{eqn17}) and (\ref{eqn18}), which form a weakened version of the axiomatization of a Boolean algebra. GKAT generalises KAT in the following sense:

\begin{lemma}\label{lemma1}
	Any KAT is a GKAT.
\end{lemma}
\proof{
	For a fixed KAT \[\boldsymbol{A}=(K,T,+,;,^*,\bar{}\ ,0,1)\] define
	\[\boldsymbol{M}=(K,T,+,;,^*,\rightarrow,0,1)\]
	inheriting the operators $+$, $;$, $^*$ and constants $0$ and $1$ from $\boldsymbol{A}$. Let $a\to b:= \bar{a}+b$, for $a,b\in T$.




	
	
	\noindent The crucial part of the proof verifies that axiom (\ref{eqn14}) holds for $\boldsymbol{M}$, for all $a,b,c\in T$.
	\noindent To see that, assume $a;b\leq c$. Then,
	\begin{eqnarray*}
		\arrayin{
			& & a;b\leq c
			\just\Leftrightarrow{$;$ is the conjunction of tests}
			a\wedge b\leq c
			\just\Leftrightarrow{commutativity of $\wedge$}
			b\wedge a\leq c
			\just\Leftrightarrow{test shunting}
			b\leq \bar{a}+c
			\just\Leftrightarrow{definition of $\to$}
			b\leq a\to c
		}
	\end{eqnarray*}

	We have just shown that axiom (\ref{eqn14}) holds for any $a,b,c\in T$ in $\boldsymbol{M}$.
	Since axioms (\ref{eqn1})-(\ref{eqn13}), (\ref{eqn17}), (\ref{eqn18}) are axioms of $\boldsymbol{A}$, $\boldsymbol{M}$ is indeed a GKAT.}
\vspace{-2mm}

\begin{example}\label{ex:2}
(\textbf{2} - the Boolean lattice).
Our first example is the well-known binary structure
\newline
\[\boldsymbol{2} = (\{\top, \bot\},\{\top, \bot\},\vee,\wedge,^*,\to,\bot,\top)\]
\newline
\noindent with the standard interpretation of Boolean connectives. Operator $^*$ maps each element of $\{\top,\bot\}$ to $\top$ and $\to$ corresponds to logical implication.
\end{example}

\begin{example}\label{ex:3}
	\noindent A second example is provided by the three-element linear lattice, which introduces an explicit denotation $u$ for ``unknown'' (or ``undefined'').
	\newline
	\[\boldsymbol{3} = (\{\top, u, \bot\},\{\top, u, \bot\},\vee,\wedge,^*,\to,\bot,\top)\]
	\newline
	where
	\newline
	\begin{table}[H]
		\fontsize{10pt}{10pt}
		\selectfont
		\centering
		\begin{tabular}{l|ccccc}
			$\vee$ & $\bot$ & u & $\top$ & \\
			\hline
			$\bot$ & $\bot$ & $u$ & $\top$ & \\
			$u$ & $u$ & $u$ & $\top$ & \\
			$\top$ & $\top$ & $\top$ & $\top$ &
		\end{tabular} \hspace{0.5 cm}
		\begin{tabular}{l|ccccc}
			$\wedge$ & $\bot$ & $u$ & $\top$ & \\
			\hline
			$\bot$ & $\bot$ & $\bot$ & $\bot$ & \\
			$u$ & $\bot$ & $u$ & $u$ & \\
			$\top$ & $\bot$ & $u$ & $\top$ &
		\end{tabular} \hspace{0.5 cm}
		\begin{tabular}{l|ccccc}
			$\to$ & $\bot$ & u & $\top$ & \\
			\hline
			$\bot$ & $\top$ & $\top$ & $\top$ & \\
			$u$ & $\bot$ & $\top$ & $\top$ & \\
			$\top$ & $\bot$ & $u$ & $\top$ &
		\end{tabular} \hspace{0.5 cm}
		\begin{tabular}{l|c}
			$^*$ & \\
			\hline
			$\bot$ & $\top$ \\
			$u$ & $\top$ \\
			$\top$ & $\top$
		\end{tabular}
	\end{table}
\end{example}

\begin{example}\label{ex:powerset}
\noindent For a fixed, finite set $A$, another instance of GKAT is
\newline
\[\boldsymbol{2^A} = (P(A),P(A),\cup,\cap,^*,\to,\emptyset,A)\]
\newline
\noindent where $P(A)$ denotes the powerset of $A$, $\cup$ and $\cap$ are set union and intersection, respectively, $^*$ maps each set $X\in P(A)$ into $A$, and $X\to Y=X^C\cup Y$, where $X^C=\{x\in A\mid x\notin X\}$. 	
\end{example}
\begin{example}\label{ex:luka}
\noindent Another example is based on the well-known \L ukasiewicz arithmetic lattice.
\newline
\[\boldsymbol{\L}=([0,1],[0,1],max,\odot,^*,\to,0,1)\]
\newline
\noindent where $x\to y=min\{1,1-x+y\}$, $x\odot y=max\{0,x+y-1\}$ and $^*$ maps each point of the interval $[0,1]$ to $1$.
\end{example}
\begin{example}\label{ex:product}
	As another example, consider the standard $\Pi$-algebra
	\newline
	\[\boldsymbol{\Pi}=([0,1],[0,1],max,.,^*,\to,0,1)\]
	\newline
	\noindent where $.$ is the usual multiplication of real numbers,
	\newline
	\begin{equation*}
		x\to y=
		\begin{cases}
		1, & \text{if}\ x\leq y \\
		y/x, & \text{if}\ y< x
		\end{cases}
	\end{equation*}
	\newline
	$ / $ is real division and $^*$ maps each point of the interval $[0,1]$ to $1$.
\end{example}

\begin{example}\label{ex:godel}
	A Gödel algebra is also an instance of GKAT. Actually,
	\newline
	\[G = ([0,1],[0,1],max,min,^*,\to,0,1)\]
	\newline
	where \begin{equation*}
	x\to y=
	\begin{cases}
	1,\ \text{if}\ x\leq y\\
	y,\ \text{if}\ y< x
	\end{cases}
	\end{equation*}
\end{example}

\noindent and $^*$ maps each point of the interval $[0,1]$ to $1$.

\begin{example}\label{ex:woop}
Let us consider now a GKAT endowing the finite \textit{Wajsberg hoop} with a star operator \cite{Hoops2000}. For a fixed natural $k$ and a generator $a$, one gets
\newline
\[\boldsymbol{W_k}=(W_k,W_k,+,;,^*,\to,0,1)\]
\newline
\noindent where $W_k=\{a^0,a^1,...,a^{k-1}\}$, $1=a^0$ and $0=a^{k-1}$. Moreover, for any $m,n\leq k-1, a^m+a^n=a^{min\{m,n\}}, a^m;a^n=a^{min\{m+n,k-1\}}, (a^m)^*=a^0$ and $a^m\to a^n=a^{max\{n-m,0\}}$.
\end{example}
\begin{example}\label{infwoop}
 The $(min, +)$ Kleene algebra \cite{DAA}, known as the tropical semiring, can be extended to a GKAT by adding residuation $\rightarrow$. First, let $R_+$ denote the set $\{x\in \mathbb{R}\mid x\geq 0\}$ and adjoin $\infty$ as a new constant. Thus, define
\newline
\[\boldsymbol{R}=(R_+\cup\{\infty\},R_+\cup\{\infty\},min,+,^*,\to,\infty,0)\]
\newline
\noindent where, for any $x,y\in R_+\cup\{\infty\}$, $x^*=0$ and $x\rightarrow y=max\{y-x,0\}$.
\end{example}

Example \ref{ex:2} represents the algebraic semantics of classical two-valued logic, while Example \ref{ex:powerset} operates over sets. To reason in discrete multi-valued logics, examples \ref{ex:3} and \ref{ex:woop} are pertinent. For the purpose of this work, i.e. for reasoning about graded computations and assertions in a multi-valued truth space, Examples \ref{ex:luka}, \ref{ex:product} and \ref{ex:godel} are particularly relevant, since they correspond to well-known models for fuzzy and multi-valued logics. 
Note that in all examples considered, $T=K$, that is, the set of tests and the set of programs coincide.

As stated above, while tests in KAT have a binary outcome, such is not necessarily the case in GKAT in which tests are graded. This entails the need to weaken the Boolean subalgebra $(T,+,;,^*,\bar{}\ ,0,1,)$ of KAT. In any GKAT, for any test $a\in T$, $a;(a\rightarrow 0) = 0$\label{eqn22} which follows immediately from definition of $\leq$ and axiom (\ref{eqn14}).
	%
		%
		%
		%
		%
However, it is not necessarily true that $a+(a\rightarrow 0) = 1$\label{eqn23}.
Let us illustrate this in the following example.

\begin{example}\label{ex:9}
	
	Consider the GKAT \[(\{0,n,m,1\}, \{0,m,1\}, +, ;, ^*, \to 0, 1)\] in which the operation $^*$ maps all points to the top element $1$, and the remaining operations are defined as follows:
	\begin{table}[H]
		\fontsize{10pt}{10pt}
		\selectfont
	\centering
\begin{tabular}{l|ccccc}
 + & $0$ & $n$ & $m$ & $1$ & \\
 \hline
$0$ & $0$ & $n$ & $m$ & $1$ & \\
$n$ & $n$ & $n$ & $m$ & $1$ & \\
$m$ & $m$ & $m$ & $m$ & $1$ & \\
$1$ & $1$ & $1$ & $1$ & $1$ & 
\end{tabular} \hspace{0.5 cm}
\begin{tabular}{l|ccccc}
 ; & $0$ & $n$ & $m$ & $1$ & \\
 \hline
$0$ & $0$ & $0$ & $0$ & $0$ & \\
$n$ & $0$ & $0$ & $0$ & $n$ & \\
$m$ & $0$ & $0$ & $0$ & $m$ & \\
$1$ & $0$ & $n$ & $m$ & $1$ & 
\end{tabular} \hspace{0.5 cm}
\begin{tabular}{l|ccccc}
 $\rightarrow$ & $0$ & $n$ & $m$ & $1$ &\\
 \hline
$0$ & $1$ & $0$ & $1$ & $1$ &\\
$n$ & $0$ & $0$ & $0$ & $0$ &\\
$m$ & $m$ & $0$ & $1$ & $1$ &\\
$1$ & $0$ & $0$ & $m$ & $1$ & 
\end{tabular}
\end{table}

\end{example}

 \noindent Clearly, $a=m$ entails $m+(m\rightarrow 0)= m+m = m \neq 1$.
It is therefore safe to state that GKAT has embedded a weakened Boolean subalgebra and, consequently, tests can assume a wider range of values,
representing the truth degree of the statement ``$b$ holds''. 
Consequently, the expression $b;p$ means that the execution of a program $p$ is guarded by that particular truth (graded) value. 


%% file: hoare.tex
%
Kleene algebra with tests provides a framework to reason about imperative programs in a (quasi) equational way. Actually, its classical presentation \cite{Kozen2000} aimed at the reduction of PHL to ordinary equations and quasi-equations, as mentioned in the introduction. In particular, the inference rules of Hoare logic are derived as theorems in KAT.

Similarly, let us explore a possible encoding of propositional Hoare logic into GKAT. Since this new structure deals with graded tests, both the meaning of Hoare triples and the inference rules need to be adjusted. This reinterpretation leads to a generalised version we shall refer to as \emph{graded propositional Hoare logic} (GPHL).

In the presence of graded tests, the interpretation of a triple $\{b\}p\{c\}$, and hence, the correctness of a program, relies on the idea that whenever $b;p$ executes with truth degree $b$, if and when it halts, it is guaranteed that $(b;p);c$ holds with at least the same degree of truth. By other words, correctness of a program can only grow with execution. Therefore, the encoding in GKAT is captured by the following inequality:
\begin{equation*}
b;p\leq b;p;c
\end{equation*}
Moreover, the equivalence
\begin{equation}
b;p \leq b;p;c\Leftrightarrow b;p = b;p;c \label{eq20},
\end{equation}
also holds in GKAT, following directly from (\ref{eqn7}), (\ref{eqn17}) and (\ref{eqn6}).
Note, however, that the equivalence 

\begin{equation*}
b;p=b;p;c\Leftrightarrow b;p\leq p;c
\end{equation*}

\noindent does not hold in GKAT.

The inference rules of Hoare logic are encoded in GKAT, as follows.
\begin{theorem}\label{theorem1}
The following implications are theorems in GKAT. 
\begin{enumerate}
\item \textit{Composition rule}:

$$b;p\leq b;p;c \wedge c;q\leq c;q;d \;\; \Rightarrow \;\; b;p;q= b;p;q;d$$
\newline
\item \textit{Conditional rule}:

$$b;c;p\leq b;c;p;d\; \wedge\; (b\rightarrow 0);c;q\leq (b\rightarrow 0);c;q;d
\;\;\Rightarrow\;$$ $$\; c;(b;p + (b\rightarrow 0);q) \leq c;(b;p + (b\rightarrow 0);q);d$$
\newline
\item \textit{Weakening and Strengthening rule}:

$$b'\leq b \wedge b;p\leq b;p;c \wedge c\leq c' \;\;\Rightarrow\;\; b';p\leq b';p;c'$$
\end{enumerate}
\end{theorem}

\proof{
See Appendix \ref{sec:th1}.}

The attentive reader certainly noticed the absence of an encoding of the While rule in the graded setting. In analogy with what was done before, such a rule would take the form:
\begin{equation}\label{eq:while}
b;c;p\leq b;c;p;c\Rightarrow c;(b;p)^*;(b\rightarrow 0)\leq c;(b;p)^*;(b\rightarrow 0);(b\rightarrow 0);c
\end{equation}
However, this is not necessarily true for all $p\in K$ and $b, c\in T$.
To see this, consider the GKAT structure of Example \ref{ex:9}. 
If $b=0, c=m, p=0$, by (\ref{eqn9}) and (\ref{eqn4}), the instantiation of $b;c;p\leq b;c;p;c$ boils down to

$$0;m;0 + 0;m;0;m = 0;m;0;m \Leftrightarrow 0=0$$

\noindent and that of $c;(b;p)^*;(b\rightarrow 0)\leq c;(b;p)^*;(b\rightarrow 0);(b\rightarrow 0);c$ becomes, by (\ref{eqn9}), 
(\ref{eqn6}) and (\ref{eqn4}),

$$m;(0)^*;1+ m;(0)^*;1;1;m =
	m;(0)^*;1;1;m\Leftrightarrow
	m = 0$$


\noindent Using these two equations, the equational implication which could represent the While rule (\ref{eq:while}), boils down to $0=0\Rightarrow m=0$, which is obviously false. The next section addresses this problem, by proposing an alternative algebraic structure able to accommodate a complete encoding of Hoare logic.


%% file: heyting.tex


By carfefully observing the encoding of the PHL while rule in KAT, it becomes apparent that one cause of failure of an analogous encoding in GKAT, mentioned in the previous section, is the impossibility of duplicating graded tests. Actually, in GKAT, $b;b=b$ does not hold, but only $b;b\leq b$ instead.
The solution proposed here is to refine the GKAT structure with some additional properties such that, i) it allows for a complete encoding of Hoare logic and, at the same time, ii) captures non-classical examples, with some degrees of uncertainty in program execution and evaluation of tests.
The idea is to resort to a stronger algebra to model the tests, instead of the Boolean algebra implicitly used in KAT.

\begin{definition}\label{def:heyting}
	 An \emph{idempotent graded Kleene algebra with tests} (I-GKAT) is a tuple
	\[(K,T,+,;,^*,\rightarrow,0,1)\]
	where $K$ and $T$ are sets, with $T\subseteq K$, $0$ and $1$ are constants in $T$, $+$ and $;$ are binary operations in both $K$ and $T$, $^*$ is a unary operator in $K$, and $\rightarrow$ is an operator only defined in $T$, satisfying the axioms in Figure \ref{fig:ax} plus the axiom below:
		
\begin{equation}\label{eq25}
	a ; a = a
\end{equation}


\end{definition}

Note that, as in GKAT, negation is not explicitly denoted, but can be derived as $a\to 0$.

\begin{figure}[ht]
	\centering
	\includegraphics[width=0.59\textwidth]{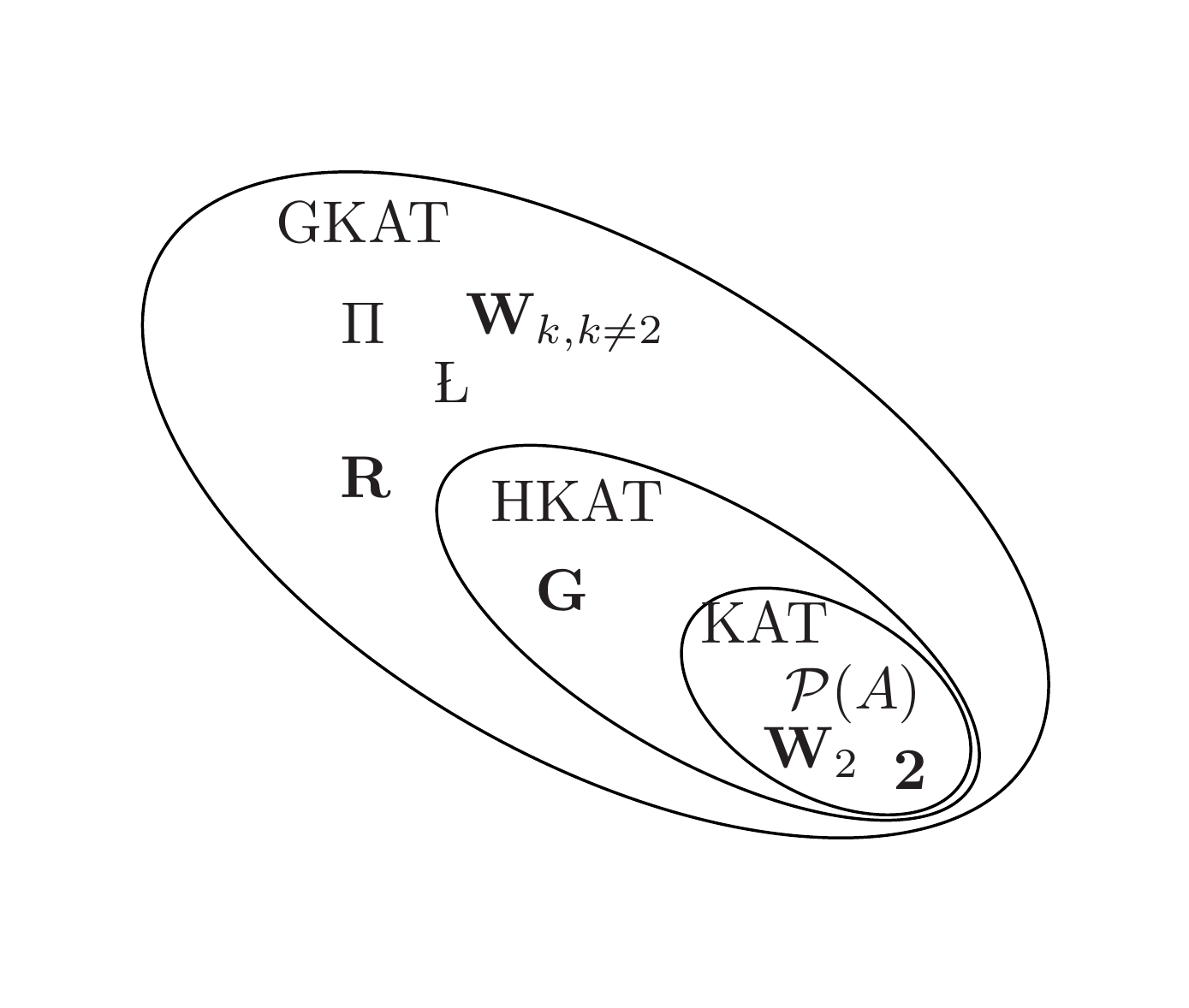}
	\caption{Examples of KAT, GKAT and I-GKAT.}
	\label{fig:diagram}
\end{figure}

The following result establishes I-GKAT as a strict subclass of GKAT, as well as another generalisation of KAT. Examples \ref{ex:2}, \ref{ex:3}, \ref{ex:powerset} and \ref{ex:godel} are instances of I-GKAT. Figure \ref{fig:diagram} sums up our results.

\begin{lemma}\label{lemma2}
	Any KAT is a I-GKAT, which in turn is also a GKAT.
\end{lemma}

\proof{It suffices to show that axiom (\ref{eqn14}) holds for all $a,b,c\in T$.
	The proof is similar to that of Lemma \ref{lemma1}.
}

I-GKAT provides a setting to discuss the behaviour of programs guarded by tests in an uncertain execution. For instance, in Example \ref{ex:3}, if $b=u$, expression $u;p$ means that one cannot be sure whether program $p$ can be executed or not.


\subsection{Propositional Hoare logic in I-GKAT}
Let us now discuss how to encode propositional Hoare logic in I-GKAT.
Differently from what happens in GKAT, the three encodings proposed by D. Kozen for Hoare logic are equivalent in I-GKAT:

\begin{equation*}
b;p=b;p;c\Leftrightarrow b;p\leq b;p;c\Leftrightarrow b;p\leq p;c
\end{equation*}

\noindent 
Hence, the inference rules of Hoare logic can be encoded in I-GKAT as they are in classical propositional Hoare logic.

\begin{theorem} The following implication is a theorem in I-GKAT.
	
	
	$$b;c;p\leq b;c;p;c\;\;\Rightarrow\;\; c;(b;p)^*;(b\rightarrow 0)\leq c;(b;p)^*;(b\rightarrow 0);(b\rightarrow 0);c$$
	
	
\end{theorem}

\proof{
	Assume, by (\ref{eqn18}),
	
	\begin{equation}\label{eq28}
		b;c;p\leq b;c;p;c\Leftrightarrow c;b;p\leq c;b;p;c
	\end{equation}

	\noindent Let us start by proving

	\begin{eqnarray*}
		 	\arrayin{
		 		& & 	c+c;(b;p)^ *;c;b;p
		 		\just\leq{by (\ref{eq28})}
		 		 c+c;(b;p)^*;c;b;p;c
		 		\just\leq{by (\ref{eq25}) and (\ref{eqn6})}
		 		c;1;c+c;(b;p)^*;c;b;p;c
		 					} \vline
		 	\arrayin{
		 		& & \just\leq{by distributivity}
		 		c;(1+(b;p)^*;c;b;p);c
		 		\just\leq{by monotonicity}
		 		c;(1+(b;p)^*;b;p);c
		 		\just\leq{by (\ref{eqn11})}
		 		c;(b;p)^*;c
		 	}
		 \end{eqnarray*}

\noindent But

\begin{eqnarray*}
	& & c+c;(b;p)^ *;c;b;p\leq c;(b;p)^*;c
	\just\Rightarrow{(\ref{eqn13})}
	c;(b;p)^*\leq c;(b;p)^*;c
	\just\Rightarrow{monotonicity of $;$}
	c;(b;p)^*;(b\to 0)\leq c;(b;p)^*;c;(b\to 0)
	\just\Leftrightarrow{(\ref{eqn18})}
	c;(b;p)^*;(b\to 0)\leq c;(b;p)^*;(b\to 0);c
	\just\Leftrightarrow{(\ref{eq25})}
	c;(b;p)^*;(b\to 0)\leq c;(b;p)^*;(b\to 0);(b\to 0);c
\end{eqnarray*}
}

	
	
	
	
	
	

%% file: fuzzy.tex

\subsection{Preliminaries}

This section illustrates both GKAT and I-GKAT constructions by discussing how they can be developed over fuzzy sets, fuzzy relations, fuzzy languages and matrices.

\begin{definition}
\label{def:fset}
	Given a set $X$ and a complete residuated lattice
	$\mathbf{W}$ over carrier $W$, a \emph{fuzzy subset} of $X$ is a function $\varphi :X\to W$; $\varphi(x)$ defines the \emph{membership degree of $x$ in $\varphi$}.
\end{definition}


%

\begin{definition}
\label{def:frel}
	Let $X_1,X_2,\ldots,X_n$ be sets. A \emph{fuzzy relation} $\mu$ between\\
	$X_1,X_2,\ldots,X_n$ is a fuzzy subset of the Cartesian product $X_1\times X_2\times\cdots\times X_n$.
\end{definition}

%
%
%

For each $x_1\in X_1,x_2\in X_2,\ldots, x_n\in X_n$, $\mu(x_1,x_2,\ldots,x_n)$ can be interpreted as the truth value of how elements $x_1,x_2,\ldots,x_n$ are related by $\mu$. Therefore, as fuzzy sets model collections of objects, fuzzy relations model relationships between objects up to some membership degree. For the purpose of this work, we consider only binary fuzzy relations. So, every time we mention the term \emph{fuzzy relation}, we are referring to fuzzy subsets of the Cartesian product $X_1\times X_2$ .

\begin{definition}
\label{def:flang}
	Let $\Sigma$ be an alphabet, $W$ the carrier of a complete residuated lattice
	 $\mathbf{W}$, and consider $\Sigma^*$ the set of words over $\Sigma$. A \emph{fuzzy language} over $\Sigma$ is a fuzzy subset of $\Sigma^*$, that is, a function $\lambda:\Sigma^*\to W$.
\end{definition}

Note that all the above concepts are defined over a complete residuated lattice. One one hand, a complete lattice is needed to guarantee the existence of suprema for all subsets of $W$. On the other hand, the residuum $\to$ is used in the context of this work to define a generalised negation for the values of $W$.
\subsection{Building GKAT and I-GKAT structures}

Consider two complete residuated lattices $\K$ and $\T$ over, respectively, carriers $K$ and $T$. Fuzzy sets, fuzzy relations and fuzzy languages may be presented as functions from their domain to, respectively, $K$ and $T$. 
We denote by $+$ the supremum of $\K$, and operators $;$ and $\to$ satisfying the axioms $(\ref{eqn1})$-$(\ref{eqn9})$ and $(\ref{eqn14})$ of Figure $\ref{fig:ax}$. We use the same notation for operators of $\T$ satisfying $(\ref{eqn1})$-$(\ref{eqn9})$ plus $(\ref{eqn14})$-$(\ref{eqn18})$. 
Since $+$ and $;$ are associative, we can generalise them to $n$-ary operators and use the notation $\sum$ and $\prod$ to represent their iterated versions, respectively.
For the specific constructions presented in this section (as given in Definitions \ref{defFSET}, \ref{defFREL} and \ref{defFLANG}), we assume 
both $\K$ and $\T$ to be complete residuated lattices, ensuring that the following properties hold:

\begin{equation}\label{eq:infdist1}
a;(\sum_{i\in I}b_i)=\sum_{i\in I}(a;b_i)
\end{equation}

\begin{equation}\label{eq:infdist2}
(\sum_{i\in I}b_i);a=\sum_{i\in I}(b_i;a)
\end{equation}
where $I$ is a (possibly infinite) index set.
To formalise these structures as I-GKAT, we consider $;$ to be also idempotent, i.e. satisfying (\ref{eq25}).



\begin{definition}\label{defFSET}

Let $X$ be a set and $\T$ a complete residuated lattice over carrier $T$. 
The algebra of \emph{fuzzy sets} over $\T$ is the structure

\[\mathbf{FSET(\T)} = (T^X,T^X,\cup,\otimes,^*,\to,\varnothing,\chi)\]

\noindent where $T^X$ is the set of all fuzzy sets over $X$ and, for all $\varphi, \psi \in T^X$ and $x \in X$, operators are defined pointwise by

\begin{eqnarray*}
	(\varphi\cup\psi)(x)&=&\varphi(x)+\psi(x)\\
	(\varphi\otimes\psi)(x)&=&\varphi(x);\psi(x)\\
	(\varphi^*)(x)&=&\sum\limits_{k\geq0}
	 \varphi^k(x)
	 \\
	(\varphi\to\psi)(x)&=&\varphi(x)\to\psi(x)\\
	\varnothing(x)&=&0\\
	\chi(x)&=&1
\end{eqnarray*}

\noindent with $\varphi^0(x)=\chi(x)$ and $\varphi^{k+1}(x)=(\varphi^n\otimes \varphi)(x)$. The values of fuzzy sets, $\varphi(x)$ and $\psi(x)$, are elements of $T$, and $0,1$ are, respectively, the least and greatest elements of $T$. 
The partial order $\subseteq$ for fuzzy sets is given by

\begin{equation*}
\varphi \subseteq \psi \Leftrightarrow \forall x\in X. \varphi(x)\leq \psi(x), \varphi,\psi\in T^X
\end{equation*}

\noindent where $\leq$ is the order of Definition 2.
\end{definition}


Note that, in this definition, the two sets of the signature of $\mathbf{FSET(\T)}$ coincide, both defined as functions with codomain $T$.

\begin{theorem}\label{theoFSET}
For any complete residuated lattice $\T$ satisfying (\ref{eqn18}), 
$\mathbf{FSET(\T)}$ forms a GKAT and. If $\T$ satisfies (\ref{eq25}) $\mathbf{FSET(\T)}$ forms a I-GKAT.
\end{theorem}

\proof{
See Appendix \ref{sec:th3}.}

A similar approach can be followed in the case of fuzzy relations. We start by defining an algebra of such relations over complete residuated lattices $\K$ and $\T$.

\begin{definition}\label{defFREL}
	
Let $X$ be a set, $\K$ and $\T$ complete residuated lattices ($\T$ satisfies (\ref{eqn18})) 
over, respectively, carriers $K$ and $T$. 
The algebra of \emph{fuzzy relations} over $\K$ and $\T$ is defined as

\[\mathbf{FREL(\K,\T)} = (K^{X\times X},T^{X\times X},\cup,\circ,^*,\to,\varnothing,\Delta)\]

\noindent where $K^{X\times X}$ is the set of all fuzzy relations over $X\times X$, the elements of $T^{X\times X}$ are diagonal fuzzy relations, i.e. fuzzy relations $\sigma$ such that $\sigma(x,y)=0$ whenever $x \neq y$. Moreover, for all $\mu, \nu \in K^{X\times X}$, $\sigma, \eta \in T^{X\times X}$, $x,y,z \in X$, the operators are defined by

\begin{eqnarray*}
	(\mu\cup\nu)(x,y)&=&\mu(x,y)+\nu(x,y)\\
	(\mu\circ\nu)(x,y)&=&\sum\limits_{z\in X} \mu(x,z);\nu(z,y)\\
	(\mu^*)(x,y)&=&\sum\limits_{k\geq 0}^{} \mu^k(x,y)
	\\
	(\sigma\to\eta)(x,y)&=&\begin{cases}
		\sigma(x,y)\to\eta(x,y), & \text{if } x=y\\
		0, & \text{otherwise}
	\end{cases}\\
	\varnothing(x,y)&=&0\\
	\Delta(x,y)&=&\begin{cases}
		1, & \text{if } x=y\\
		0, & \text{otherwise}
	\end{cases}
\end{eqnarray*}

\noindent with $\mu^0(x,y)=\Delta(x,y)$, $\mu^{k+1}(x,y)=(\mu^{k}\circ \mu)(x,y)$. The values of fuzzy relations, $\mu(x,y)$ and $\nu(x,y)$, are elements of $K$, the values of $\sigma(x,y)$ and $\eta(x,y)$ are elements of $T$, and, finally, constants $0,1$ are the least and the greatest elements of $T$. 
Similarly to the previous one, the partial order $\subseteq$ for fuzzy relations is given by

\begin{equation*}
\mu \subseteq \nu \Leftrightarrow \forall (x,y)\in X\times X. \mu(x,y)\leq \nu(x,y), \mu,\nu\in K^X
\end{equation*}

\noindent where $\leq$ is the order referred in Definition 2.

\end{definition}


\begin{theorem}\label{theoFREL}
	Given 
	complete residuated lattices $\K$ and $\T$ ($\T$ satisfies (\ref{eqn18}))
	, $\mathbf{FREL(\K, \T)}$ is a GKAT. If $\T$ satisfies (\ref{eq25}) $\mathbf{FREL(\K, \T)}$ is also a I-GKAT.
\end{theorem}

\proof{
	See Appendix \ref{sec:th4}.}

\begin{definition}\label{defFLANG}
	Let $\Sigma$ be an alphabet, $\Sigma^*$ the set of all words over $\Sigma$ and $\K$, $\T$ complete residuated lattices ($\T$ satisfies (\ref{eq25})).
	The algebra of fuzzy languages over $\K$, $\T$ is defined as
	
	\[\mathbf{FLANG(\K,\T)}=(K^{\Sigma^*},T^{\Sigma^*},\cup,\cdot,^*,\to,\varnothing,\epsilon)\]	
	\noindent where $K^{\Sigma^*}$ stands for the set of all fuzzy languages over $\Sigma$, the elements of $T^{\Sigma^*}$ are languages defined by	
	\\
	
	$\iota(a_1\ldots a_n)=\begin{cases}
	a, & \text{if } a_1\ldots a_n=\epsilon, \text{ with } \epsilon \text{ being the empty word}\\
	0, & \text{otherwise}
	\end{cases}$ \\where $a\in T$
	\\
	
	\noindent and, for all $\lambda_1,\lambda_2\in K^{\Sigma^*}$ and all $\iota_1,\iota_2\in T^{\Sigma^*}$, given a word $a_1\dots a_n\in \Sigma^*$, the operators $\cup$, $\cdot$, $^*$, $\to$, $\varnothing$ and $\epsilon$ are defined as:
	
	\begin{eqnarray*}
		(\lambda_1\cup\lambda_2)(a_1\ldots a_n)&=&\lambda_1(a_1\ldots a_n)+\lambda_2(a_1\ldots a_n)\\
		(\lambda_1\cdot \lambda_2)(a_1\ldots a_n)&=&\sum_{i=1}^{n-1}\lambda_1(a_1\ldots a_i);\lambda_2(a_{i+1}\ldots a_n)\\
		(\lambda^*)(a_1\ldots a_n)&=&\sum\limits_{k\geq0}
		\lambda^k(a_1\ldots a_n)\\
		(\iota_1\to\iota_2)(a_i\ldots a_n)&=&\begin{cases}
	\prod_{a_1\ldots a_{i-1}}^{}\big(\iota_1(a_1\ldots a_{i-1})\to\iota_2(a_1\ldots a_n)\big), i\leq n, \\ \text{ if }a_1\ldots a_{i-1}=\epsilon\\
	0, \text{ otherwise}
		\end{cases}\\
		\varnothing(a_1\ldots a_n)&=&0\\
		\epsilon(a_1\ldots a_n)&=&\begin{cases}
			1 & \text{if } a_1\dots a_n=\epsilon \text{, with $\epsilon$ being the empty word}\\			0 & \text{otherwise}
		\end{cases}
	\end{eqnarray*}
	
	\noindent with $\lambda^0(a_1\ldots a_n)=\epsilon(a_1\ldots a_n)$ and $\lambda^{k+1}(a_1\ldots a_n)=(\lambda^k.\lambda)(a_1\ldots a_n)$. The values of fuzzy sets, $\lambda_1(a_1\ldots a_n)$ and $\lambda_2(a_1\ldots a_n)$, are elements of $K$, and $0,1$ are the least and greatest elements of $T$.
	The partial order $\subseteq$ for fuzzy languages is given by
	
	\begin{equation*}
	\lambda_1 \subseteq \lambda_2 \Leftrightarrow \forall a_1\ldots a_n\in \Sigma^*. \lambda_1(a_1\ldots a_n)\leq \lambda_2(a_1\ldots a_n), \lambda_1,\lambda_2\in K^{\Sigma^*}
	\end{equation*}
	
	\noindent where $\leq$ is the order of Definition 2.
	
\end{definition}

\begin{theorem}\label{theoFLANG}
	
Given complete residuated lattices
$\K$ and $\T$ ($\T$ satisfies (\ref{eqn18}) and (\ref{eq25})), $\mathbf{FLANG(\K, \T)}$ is a I-GKAT.
	
\end{theorem}

\proof{
	Since a fuzzy language $\lambda$ is a fuzzy subset of a set of elements (in this case, the alphabet $\Sigma^*$) and the operators $\cup$ and $^*$ are defined as $\cup$ and $^*$ in $\mathbf{FSET(\T)}$, respectively, and $\cdot$ as $\circ$ in $\mathbf{FREL(\K,\T)}$, the proof is identical to that of Theorem \ref{theoFSET} for operators $\cup$ and $^*$, and to that of Theorem \ref{theoFREL} for the $\cdot$ operator.
	
	It remains to prove axiom (\ref{eqn14}):
Take $\iota_1,\iota_2,\iota_3\in T^{\Sigma^*}$ and $v\in\Sigma^*$. Consider first the case $v\neq\epsilon$.
	
	\noindent Assuming $(\iota_1\cdot\iota_2)(v)\leq\iota_3(v)\Leftrightarrow\sum_{v_1,v_2}\iota_1(v_1);\iota_2(v_2)\leq\iota_3(v)\Leftrightarrow 
	\iota_3(v)=0$ we want to prove that $
	\iota_2(v)\leq(\iota_1\to\iota_3)(v) $. But, by definition of $\iota$ and $\to$, $\iota_2(v)\leq(\iota_1\to\iota_3)(v)\Leftrightarrow 0\leq 0$.
	
	\noindent Consider now $v=\epsilon$. 
	We want to prove that
	
	$$(\iota_1\cdot\iota_2)(\epsilon)\leq\iota_3(\epsilon)\Leftrightarrow\iota_2(\epsilon)\leq(\iota_1\to\iota_3)(\epsilon)$$
	
	
	
	\begin{eqnarray*}
		\arrayin{
			& & \iota_2(\epsilon)\leq(\iota_1\to\iota_3)(\epsilon)
			\just\Leftrightarrow{definition of $\to$}
			\iota_2(\epsilon)\leq\prod_u(\iota_1(u)\to\iota_3(u\epsilon))
			\just\Leftrightarrow{definition of $\prod$}
			\iota_2(\epsilon)\leq(\iota_1(u_1)\to\iota_3(u_1\epsilon));\ldots;(\iota_1(u_{n-1})\to\iota_3(u_{n-1} \epsilon));(\iota_1(\epsilon)\to\iota_3(\epsilon \epsilon)),\\ & & u_1,\ldots, u_{n-1}\neq \epsilon
			\just\Leftrightarrow{definition of $\iota$}
			\iota_2(\epsilon)\leq(0\to 0);\ldots;(0\to 0);(\iota_1(\epsilon)\to \iota_3(\epsilon))
			\just\Leftrightarrow{$0\to0=1$ for all integral lattices (\cite{Madeira2016}) and (\ref{eqn6})}
			\iota_2(\epsilon)\leq\iota_1(\epsilon)\to \iota_3(\epsilon)
			\just\Leftrightarrow{(\ref{eqn14})}
			\iota_1(\epsilon);\iota_2(\epsilon)\leq\iota_3(\epsilon)
			\just\Leftrightarrow{definition of $\cdot$}
			(\iota_1\cdot\iota_2)(\epsilon)\leq\iota_3(\epsilon)
			}
	\end{eqnarray*}}

%% file: matrices.tex
Now we present the construction of matrices over a GKAT and I-GKAT.
\begin{definition}\label{defMAT}
Let $\boldsymbol{A}=(K,T,+,;,^*,\rightarrow,0,1)$ be a GKAT (or a I-GKAT). The algebra consisting of the family $M(n,K)$ of $n\times n$ matrices over $\boldsymbol{A}$ is defined as

\[\boldsymbol{M(n,A)}=(M(n,K),\Delta(n,T),+,;,^*,\rightarrow,0_n,I_n)\]

\noindent where $+$ and $;$ stand for the usual matrix addition and multiplication, respectively; $0_n$ is the $n\times n$ matrix of zeros and $I_n$ the $n\times n$ identity matrix.
The subalgebra is the set $\Delta(n,T)$ of $n\times n$ diagonal matrices, with operators $+$ and $;$ and matrices $0_n$ and $I_n$ defined in the same way. The entries of the diagonal matrices are elements of the subalgebra $(T,+,;,0,1)$ of GKAT (or I-GKAT) $\A$. Finally, operation $\rightarrow$ is defined as:

\begin{center}
	$A\rightarrow B=\begin{cases}
	A_{ij}\rightarrow B_{ij} & \text{if } i=j\\
	0 & \text{otherwise}
	\end{cases}$
\end{center}

\end{definition}

\begin{theorem}\label{theoMAT}
$\boldsymbol{M(n,A)}$ is a GKAT and a I-GKAT.
	
\end{theorem}

\proof{
See Appendix \ref{sec:th6}.}

%% file: folk.tex

This section illustrates our constructions revisiting a result on denesting two nested \textbf{while} loops \cite{Kozen1997}, in a scenario where both assertions and computations are expressed in a weighted context. 
Most proofs in D. Kozen's paper \cite{Kozen1997} rely on the use of a commutativity condition ($b;p=p;b$) which asserts that the execution of program $p$ does not modify the value of test $b$. In KAT, it is possible to argue, as well, that if $p$ does not affect $b$, neither should it affect $\bar{b}$, which is formally stated through the following lemma:

\begin{lemma}\label{lem:lemma3}
	In any Kleene algebra with tests the following are equivalent:
	
	\begin{enumerate}
		\item [] (1) $b;p=p;b$
		\item [] (2) $\bar{b};p=p;\bar{b}$
		\item [] (3) $b;p;\bar{b}+\bar{b};p;b=0$
	\end{enumerate}
	
\end{lemma}

 \noindent In GKAT, however, negation is relaxed and expressed as $a\to 0$, for all $a\in T$. So, the conditions above must be written as
 
 	\begin{enumerate}
 		\item [] (1) $b;p=p;b$
 		\item [] (2) $(b\to 0);p=p;(b\to 0)$
 		\item [] (3) $b;p;(b\to 0)+(b\to 0);p;b=0$
 	\end{enumerate}

\noindent
However, it is important to note that not all implications hold in GKAT. 

\begin{lemma}\label{lem:lemma4}
$(1)\Leftrightarrow(2)$ does not hold in GKAT.
\end{lemma}

\proof{
This can be shown by the following counter example: a GKAT over the set $\{0,n,m,1\}$, with $\{0,m,1\}\subseteq T$ and $n\in K$, in which the operator $^*$ maps all points to the top element $1$ and the remaining operators are defined as follows:

\begin{table}[H]
	\fontsize{10pt}{10pt}
	\selectfont
	\centering
	\begin{tabular}{l|ccccc}
		+ & $0$ & $n$ & $m$ & $1$ & \\
		\hline
		$0$ & $0$ & $n$ & $m$ & $1$ & \\
		$n$ & $n$ & $n$ & $m$ & $1$ & \\
		$m$ & $m$ & $m$ & $m$ & $1$ & \\
		$1$ & $1$ & $1$ & $1$ & $1$ &
	\end{tabular} \hspace{0.5 cm}
	\begin{tabular}{l|ccccc}
		; & $0$ & $n$ & $m$ & $1$ & \\
		\hline
		$0$ & $0$ & $0$ & $0$ & $0$ & \\
		$n$ & $0$ & $0$ & $0$ & $n$ & \\
		$m$ & $0$ & $n$ & $m$ & $m$ & \\
		$1$ & $0$ & $n$ & $m$ & $1$ & 
	\end{tabular} \hspace{0.5cm}
	\begin{tabular}{l|ccccc}
		$\rightarrow$ & $0$ & $n$ & $m$ & $1$ & \\
		\hline
		$0$ & $1$ & $0$ & $1$ & $1$ & \\
		$n$ & $0$ & $0$ & $0$ & $0$ & \\
		$m$ & $0$ & $0$ & $1$ & $1$ & \\
		$1$ & $0$ & $0$ & $m$ & $1$ & 
	\end{tabular}
\end{table}

\noindent If $b=n, p=m$, the instantiation of $b;p=p;b\Leftrightarrow (b\to 0);p=p;(b\to 0)$ becomes

\begin{equation*}
	n;m=m;n\Leftrightarrow (n\to 0);m=m;(n\to 0)
\end{equation*}

\noindent Thus, the expression turns into $0=n\Leftrightarrow 0=0$, which is clearly false.
}


\begin{lemma}\label{lem:lemma5}
	Implications $(1)\Rightarrow(3)$ and $(2)\Rightarrow(3)$ hold in GKAT.
\end{lemma}

\proof{
Both implications arise by commutativity and the fact that $a;(a\to 0)=0$, for all $a\in T$.
}

\begin{lemma}\label{lem:lemma6}
	Implication $(3)\Rightarrow(1)$ does not hold in GKAT.
\end{lemma}

\proof{
This can be shown by the following counter example: a GKAT over the set $\{0,n,m,1\}$, with $\{0,n,1\}\subseteq T$ and $m\in K$, in which the operator $^*$ maps all points to the top element $1$ and the remaining operators are defined as follows:

\begin{table}[H]
	\fontsize{10pt}{10pt}
	\selectfont
	\centering
	\begin{tabular}{l|ccccc}
		+ & $0$ & $n$ & $m$ & $1$ & \\
		\hline
		$0$ & $0$ & $n$ & $m$ & $1$ & \\
		$n$ & $n$ & $n$ & $m$ & $1$ & \\
		$m$ & $m$ & $m$ & $m$ & $1$ & \\
		$1$ & $1$ & $1$ & $1$ & $1$ &
	\end{tabular} \hspace{0.5 cm}
	\begin{tabular}{l|ccccc}
		; & $0$ & $n$ & $m$ & $1$ & \\
		\hline
		$0$ & $0$ & $0$ & $0$ & $0$ & \\
		$n$ & $0$ & $0$ & $n$ & $n$ & \\
		$m$ & $0$ & $0$ & $m$ & $m$ & \\
		$1$ & $0$ & $n$ & $m$ & $1$ & 
	\end{tabular} \hspace{0.5cm}
	\begin{tabular}{l|ccccc}
		$\rightarrow$ & $0$ & $n$ & $m$ & $1$ & \\
		\hline
		$0$ & $1$ & $1$ & $0$ & $1$ & \\
		$n$ & $n$ & $1$ & $0$ & $1$ & \\
		$m$ & $0$ & $0$ & $0$ & $0$ & \\
		$1$ & $0$ & $n$ & $0$ & $1$ & 
	\end{tabular}
\end{table}

\noindent If $b=n, p=m$, the instantiation of $b;p;(b\to 0)+(b\to 0);p;b=0\Rightarrow b;p=p;b$ becomes $0=0\Rightarrow n=0$ which is obviously false.
}

\begin{lemma}\label{lem:lemma7}
	Implication $(3)\Rightarrow(2)$ does not hold in GKAT.
\end{lemma}

\proof{
	Consequence of Lemma \ref{lem:lemma4} and Lemma \ref{lem:lemma5}.}



The intuitive interpretation of these implications is that if $p$ preserves $b$ (or $b\to 0$), the execution of $p$ between testing $b$ and its complement, no matter which test is performed first, always halt. A similar result holds for I-GKAT and is proved along similar lines.

%
%
%

We can therefore argue that this dependency on commutativity conditions becomes a hindrance for proving most of the results on program equivalence that we intend: it is impossible to handle such proofs in a (quasi) equational way without considering them. However, the result that is, perhaps, the most interesting one, of denesting two nested \textbf{while} loops, does not resort to the commutativity conditions. 
Let us detail this example.

%% file: loops.tex


%
%


The original proof in the above mentioned paper \cite{Kozen1997} relies on one of De Morgan laws to prove the intended result. More precisely, the proof uses the rule
\[\neg( a \vee b) \equiv \neg a \wedge \neg b\] 
that can be formalised in our setting as
	\begin{equation}
		(a+b)\to 0 = (a\to 0);(b\to 0)\label{eq:demorgan}
	\end{equation}
Since, in general, this rule does not hold in I-GKAT, we have to impose it in the following characterisation. Note that the rule holds in all instances of I-GKAT enumerated in the paper, namely \ref{ex:2}, \ref{ex:3}, \ref{ex:powerset} and \ref{ex:godel}.


We are now in conditions to show that a pair of \textbf{while} loops can be transformed into a single \textbf{while} loop inside a conditional test, as formalised in the following theorem:

\begin{theorem}\label{theo:nestedwhile}
	The program
	\begin{align}
	\textbf{ while } b \textbf{ do } \textbf{ begin } \nonumber\\
	p; \nonumber\\
	& \textbf{ while } c \textbf{ do } q \label{eq29}\\
	\textbf{ end} \nonumber
	\end{align}
	
	\noindent is equivalent to
	
	\begin{align}
	\textbf{if } b \textbf{ then begin} \nonumber\\
	& p; \nonumber\\
	& \textbf{while } b+c \textbf{ do} \label{eq30}\\
	& \textbf{ if } c \textbf{ then } q \textbf{ else } p \nonumber\\
	\textbf{end} \nonumber
	\end{align}
	
\noindent in I-GKAT extended with (\ref{eq:demorgan}).
	
\end{theorem}

\proof{
The proof uses an analogous reasoning of the one presented in \cite{Kozen1997}.
}

%% file: conclusions.tex

This paper aimed at generalising Kleene algebra with tests, to reason equationally about graded computations and assertions about them evaluated in a multi-valued truth space. The propositional fragment of classic Hoare logic was revisited in this context. We also presented four algebraic constructions as models of both generalizations of Kleene algebras introduced in the paper (GKAT and I-GKAT): the set of all fuzzy sets, the set of all fuzzy relations, the set of all fuzzy languages and the family of square matrices. Finally, we discussed (quasi) equational proofs of some classical results on program equivalence in a weighted context.

A similar roadmap is followed by R. Qiao \emph{et al.} \cite{Qiao2008} leading to the introduction of a complete theory of probabilistic KAT to deal with regular programs with probabilities. However, instead of focusing on broadening the possible range of values for tests, or on adding an uncertainty concretisation to them as an immediate consequence on program execution, the authors opted to add a new operator $+_\alpha$ to the algebraic structure, where $\alpha$ is a probability. Thus, in their work, a \emph{probabilistic Kleene algebra with Tests} is defined as \[(K,T,+,+_\alpha,\cdot,^*,0,1,\bar{}\ )\] where expression $p+_\alpha q$ represents the probabilistic choice between executing a program $p$ with probability $\alpha$ or a program $q$ with probability $1-\alpha$. Other references \cite{probProgUsingHoareL,probKAprotocol} follow a similar approach introducing probabilities at the syntactic level, namely through a new choice operator. Our approach, on the other hand, opted by redefining the notions of test and program execution.
Such approach, which describes the behaviour of the probabilistic phenomena, always enforces the production of an outcome (as expressed by the requirement that the outgoing probabilities
always sum 1). Such is not the case in the framework adopted in this paper.

The idempotent variant presented in our work is related with the one based on Heyting domain semirings \cite{DesharnaisS08},
obtained by relaxing the test algebra of Kleene algebras with domain.
One difference between this structure and our approach lies on the construction of the structure itself:
while ours is purely propositional and based on KAT, the one of \cite{DesharnaisS08} makes use of a unary operator, the domain operator,
to axiomatise the test algebra, resulting in a one sorted structure. The relaxation of the test algebra is accomplished by adding an operator
on domain elements satisfying (\ref{eqn14}), with a negation defined as $p\to 0$. It would be pertinent to do a more detailed analysis
about the set of properties that can be derived for each structure.
Moreover, the authors in \cite{DesharnaisS08} point for future work a more in depth
exploration of possible applications and directions that the flexibility of the adopted method can bring.
The formalisation and the proof of the soundness of the Hoare logic deductive system using the structure based on Heyting domain semirings,
by comparing with the our approach, seems also an appropriate discussion to be made in future work. Additionally, a more recent work \cite{DesharnaisM14}
investigates a generalisation of these domain algebras to support fuzzy relations, taken as functions from pairs of elements to the
interval $[0,1]$. Different from our approach, the authors study an axiomatisation of domain and codomain operators in the setting of
idempotent left semirings, which do not require left distributivity of multiplication over addition and right annihilation of $0$. 
Note that we started this work by adopting a presentation similar to KAT when relaxing its Boolean subalgebra to obtain GKAT and I-GKAT.
In order present a clearer comparison between these structures, either axiomatically and in terms of
obtained results, we follow the same propositional presentation, based on KAT.

The approach taken in this paper, adding a residual as a logical implication to capture a multi-valued setting, is based on previous work \cite{Madeira2016}, where an action lattice is adopted as the basic algebraic structure to generate many-valued dynamic logics.
Originally derived from \emph{action algebras} \cite{onactionalgebras}, an action lattice entails both a generic space of computations, with choice, composition and iteration, and, supported by residuation, a proper truth space for a non bivalent interpretation of assertions (as a residuated lattice).
V. Pratt thought about residuation as a pure technicality to obtain a finitely-based equational variety \cite{actionlogic}.
Subsequently, the work of D. Kozen \cite{onactionalgebras} extended this notion by adding and axiomatizing a meet operation, in order to recover the closure under matricial formation typical of Kleene algebra \cite{conway1971}.


The attentive reader may wonder why concrete illustrations of the proposed formalism seem to be lacking.
Note, however, that programs are interpreted here as weighted relations and tests as truth degrees. 
Hence, as it happens in propositional Hoare logic derived from standard KAT, there is no first-order structure to interpret program variables. Consequently, there is no assignment rule neither for GPHL nor for HPHL, as presented here. Extending the formalism in this direction, in order to deal with imperative fuzzy programs is, naturally, in our agenda.

Fuzzy Arden Syntax (FAS) \cite{VETTERLEIN20101} is a fuzzy programming language designed for the medical domain, which 
extends Arden Syntax (AS) to cater for vague or uncertain information often arising in clinical situations. Due to the intuitiveness of its syntax, very close to natural language, AS and FAS are commonly used as syntax for knowledge base components in medical decision support systems \cite{STARREN1994411,ANAND2015,SAMWALD2012711}.
	
Built on the theory of fuzzy sets \cite{ZADEH1965338}, data types in FAS have been generalised to represent truth values between the extremes false and true. Moreover, the operations on these types were generalised accordingly. A particular consequence that emerges from the nature of these generalisations concerns the behaviour of conditional statements: while conditions evaluated in a bivalent truth space entails the execution of only one branch, in FAS an $\mathbf{if-then-else}$ statement may split. In such cases the variables are duplicated and both branches are executed in parallel, each with an associated truth degree. The notion of parallelism inherent to these statements leads us to rethink the behaviour of PHL variants introduced in this work: the conditional statements encoded in sections \ref{sec:graded_kleene_algebra_with_tests} and \ref{sec:heyting} illustrate non deterministic choice, despite the possible weighted nature of both computations and conditions. Indeed, the $+$ operator of Kleene algebra, used to encode the $\mathbf{if-then-else}$ statements in both GKAT and I-GKAT could be interpreted as (fuzzy) set union in all the examples listed.
	
Conditionals in FAS are an interesting case-study for the development of an algebraic formalism to specify the behaviour of conditional statements in fuzzy programming languages. 
For that an extension of the algebras presented in this work, with an appropriate operator to formalize parallelism, is currently being developed by the authors. In this setting, the works on Concurrent Kleene algebra \cite{CKATjournal} and Synchronous Kleene algebra \cite{Prisacariu10} are worth revisiting.



 


The results reported in Section \ref{sec:folk} lead us to discuss further why some properties fail in GKAT. In particular, why the 
preservation of the value of $b$ along a computation $p$ entails the corresponding preservation of $\bar{b}$ does not hold in either GKAT or I-GKAT, as it does in KAT.
 We observed, in Section \ref{sec:graded_kleene_algebra_with_tests}, that the negation operator must be relaxed in order to support a non bivalent truth space for assertions. Actually, this has influence on the validity of the properties in which it is involved. With this modification, some classical properties are lost. In particular, the law of excluded middle, necessary to prove the discussed implications, is no longer valid.


In all variants of dynamic logic discussed in the literature, even when some forms of structured computations are taken into consideration, the validity of assertions (for example, of Hoare triples annotating a program) is always stated in classical terms. This means that, even when the object of reasoning is e.g. a fuzzy program or a quantum system, the validity of an assertion over it is discussed in classical, two-valued logic.


In this work we assumed, as in classical PHL, that a Hoare triple is valid if $b;p=b;p;c$. In GKAT, this expression states that, after the execution of $p$ guarded by the truth degree of condition $b$, a state is reached where the truth degree of the post condition does not modify the value of the execution. In I-GKAT, for the case considered in example \ref{ex:3}, the variation from the classical case comes when $b=u$. Thus, the expression $b;p$ can be interpreted as ``not sure if program $p$ can be executed''. Due to the nature of the expression (an equality relation), this is clearly tied to the classical, two-valued logic: despite the graded nature of the computations, their correctness is evaluated in a bivalent truth space.
	
	
This limitation motivates an alternative approach currently under investigation. The intention is to go a step further, resorting to the same algebraic structure used to specify the computational paradigm, to give semantics to the logic used to reason about it. This will allow to discuss the validity of an assertion over a fuzzy or a quantum program in terms of a logic capturing itself fuzzy or quantum reasoning, respectively.

%% file: appendices.tex

\section{Proof of Theorem 1}\label{sec:th1}

\begin{description}
\item [1. Composition rule:] Let us assume that $b;p\leq b;p;c$ and $c;q\leq c;q;d$. By (\ref{eq20}), these inequalities are equivalent to $b;p=b;p;c$ and $c;q=c;q;d$, respectively. So, we have
\begin{eqnarray*}
	\arrayin{
	& & b;p;q
	\just = {$b;p=b;p;c$}
	b;p;c;q
	}
	\vline
	\arrayin{
	& & \just = {$c;q=c;q;d$}
	b;p;c;q;d
	\just = {$b;p=b;p;c$}
	b;p;q;d
	}	
\end{eqnarray*}
\item [2. Conditional rule:] Assume $b;c;p\leq b;c;p;d$ and $(b\rightarrow 0);c;q\leq (b\rightarrow 0);c;q;d$. First of all, observe that, for any $p,q,r,s\in K$
\begin{equation}
p \leq q \; \wedge \; r\leq s \Rightarrow p+r \leq q+s \label{eq21}
\end{equation}
\noindent To prove this, assume that $p\leq q$ and $r\leq s$, i.e. $p+q= q$ and $r+s=s$. Then, by (\ref{eqn1}) and (\ref{eqn2}), $(p+r)+(q+s) = (p+q)+(r+s) =q+s$. 
\noindent So, by (\ref{eq21}),
\begin{eqnarray*}
\arrayin{
& & b;c;p + (b\rightarrow 0);c;q \leq b;c;p;d + (b\rightarrow 0);c;q;d.
\just \Leftrightarrow {(\ref{eqn18}), (\ref{eqn7}) and (\ref{eqn8})}
c;(b;p + (b\rightarrow 0);q) \leq c;(b;p + (b\rightarrow 0);q);d
}
\end{eqnarray*}
\item [3. Weakening and Strengthening rule:] Finally, observe that, for all $b, c\in T$ and $p\in K$,
\begin{equation}
b;p\leq b;p;c \Rightarrow b;p;(c\rightarrow 0)\leq 0 \label{eq22} 
\end{equation}
Using (\ref{eq20}) to rewrite (\ref{eq22}) as
\begin{equation}
b;p= b;p;c \Rightarrow b;p;(c\rightarrow 0)= 0 \label{eq23}
\end{equation}
and, assuming $b;p= b;p;c$, we have
\begin{eqnarray*}
\arrayin{
& & b;p;(c\rightarrow 0)
\just = {$b;p= b;p;c$ assumption}
b;p;c;(c\rightarrow 0)
\just = {$a;(a\rightarrow 0) = 0$) and (\ref{eqn9})}
0
}
\end{eqnarray*}
Using (\ref{eq23}), the Weakening and Strengthening rule can be rewritten as 
\begin{equation*}
a\leq b \wedge b;p;(c\to 0)= 0 \wedge (d\to 0) \leq (c\to 0) \Rightarrow a;p;(d\to 0)= 0
\end{equation*}
which follows from the monotonicity of ``$;$''.
\end{description}

\section{Proof of Theorem 3}\label{sec:th3}

	Considering the way that the elements of $T^X$ and the operators $\cup$, $\otimes$ and $\rightarrow$ are defined, it is straightforward to verify that axioms (\ref{eqn1}) to (\ref{eqn9}) and (\ref{eqn14}) to (\ref{eqn18}), plus (\ref{eq25}) for I-GKAT, are satisfied. We prove that axioms dealing with operator $^*$ \big((\ref{eqn10}), (\ref{eqn12})\big) hold as well. Axiom (\ref{eqn13}) 
	can be proved analogously to (\ref{eqn12}).
	~\\
	
	\noindent
	\underline{Axiom (\ref{eqn10})}:
	\begin{eqnarray*}
		\arrayin{
			& & (\chi\cup(\varphi\otimes\varphi^*))(x)
			\just={definition of $T^X$}
			\chi(x)+\varphi(x);\varphi^*(x)
			\just={definition of $\varphi^*(x)$}
			\chi(x)+\varphi(x);(\sum_{k\geq0} \varphi^k(x))
			\just={definition of $\sum$}}
			\vline
			\arrayin{
			& &
			\begin{aligned}
				\chi(x)+\varphi(x);(\varphi^0(x)+\varphi^1(x)+\cdots)
			\end{aligned}
			\just={(\ref{eq:infdist1})}
			\begin{aligned}
			\chi(x)+\varphi(x);\varphi^0(x)+\varphi(x);\varphi^1(x)
			+\cdots
			\end{aligned}
			\just={definition of $\varphi^{k+1}(x)$}
			\chi(x)+\varphi(x)+\varphi^2(x)+\cdots
			\just={definition of $\sum$}
			\varphi^*(x)	
		}
	\end{eqnarray*}
	
	\noindent
	\underline{Axiom (\ref{eqn12})}\\
	Let us assume $(\varphi\otimes\psi)(x)\leq\psi(x)$, i.e. $\varphi(x);\psi(x)\leq\psi(x)$, by definition of the operators on fuzzy sets. Moreover,	
%
	\begin{eqnarray*}
		\arrayin{
			& & (\varphi^*\otimes\psi)(x)
			\just={definitions of $^*$ and $\otimes$}
			(\sum_{k\geq 0}\varphi^k(x));\psi(x)
			\just={definition of $\sum$}
			(\varphi^0(x)+\varphi^1(x)+\cdots);\psi(x)
		}
			\vline
			\arrayin{
			\just={(\ref{eq:infdist2}) and (\ref{eqn6})}
			\psi(x)+\varphi(x);\psi(x)+\cdots
			}
	\end{eqnarray*}
	By hypothesis and given that $\varphi(x);\varphi(x)\leq \varphi(x)$, for all $\varphi(x)\in T$, we conclude that
	
	$$\psi(x)+\varphi(x);\psi(x)+\cdots\leq\psi(x)$$
	
	\medskip
	\hfill $\square$

\section{Proof of Theorem 4}\label{sec:th4}

	The validity of (\ref{eqn1}) and (\ref{eqn2}) follows immediately from the definitions of operators on fuzzy relations. Let $\mu$, $\nu$, $\xi$ $\in K^{X\times X}$ and $x,y,z,w\in X$.
	~\\
	
	\noindent
	\underline{Axiom (\ref{eqn5})}:
	\begin{eqnarray*}
		& & ((\mu\circ\nu)\circ\xi)(x,y)
		\just={definition of $\circ$}
		\sum_{z\in X}(\sum_{w\in X}(\mu(x,w);\nu(w,z));\xi(z,y)
		\just={definition of $\sum$ and $z_i,w_i\in X, 1\leq i\leq n$}
		\begin{aligned}
		(\mu(x,w_1);\nu(w_1,z_1)+\cdots+\mu(x,w_n);\nu(w_n,z_1));\xi(z_1,y)+\cdots\\
		+(\mu(x,w_1);\nu(w_1,z_n)+\cdots+\mu(x,w_n);\nu(w_n,z_n));\xi(z_n,y)
		\end{aligned}
		\just={(\ref{eq:infdist2}) and (\ref{eqn5})
		}
		\begin{aligned}
		\mu(x,w_1);(\nu(w_1,z_1);\xi(z_1,y))+\cdots+\mu(x,w_n);(\nu(w_n,z_1);\xi(z_1,y))+\cdots\\
		+\mu(x,w_1);(\nu(w_1,z_n);\xi(z_n,y))+\cdots+\mu(x,w_n);(\nu(w_n,z_n);\xi(z_n,y))
		\end{aligned}
		\just={(\ref{eqn2}) and (\ref{eq:infdist1})
		}
		\begin{aligned}
			\mu(x,w_1);(\nu(w_1,z_1);\xi(z_1,y)+\cdots+\nu(w_1,z_n);\xi(z_n,y))+\cdots\\
			+\mu(x,w_n);(\nu(w_n,z_1);\xi(z_1,y)+\cdots+\nu(w_n,z_n);\xi(z_n,y))
		\end{aligned}
		\just={definition of $\sum$}
		\sum_{w\in X}(\mu(x,w);(\sum_{z\in X}(\nu(w,z);\xi(z,y))))
		\just={definition of $\circ$}
		(\mu\circ(\nu\circ\xi))(x,y)
	\end{eqnarray*}
	
	\noindent
	\underline{Axiom (\ref{eqn6})}:	
	\begin{eqnarray*}
		\arrayin{
			& & (\mu\circ\Delta)(x,y)
			\just={definition of $\circ$}
			\sum_{z\in X}\mu(x,z);\Delta(z,y)
			\just={definition of $\sum$ and $z_i\in X, 1\leq i\leq n$}
			\mu(x,z_1);\Delta(z_1,y)+\cdots+\mu(x,z_n);\Delta(z_n,y)
			\just={definition of $\Delta$}
			\begin{aligned}
			\mu(x,z_1);1+\cdots+\mu(x,z_n);1,\\
			\text{for all } \Delta(z_i,y)=1, 1\leq i\leq n
			\end{aligned}
			\just={(\ref{eqn6})
			}
			\mu(x,z_1)+\cdots+\mu(x,z_n)
			\just={definition of $\mu$}
			\mu(x,y)
			}
	\end{eqnarray*}
	
	\noindent
	\underline{Axiom (\ref{eqn7})}:	
	\begin{eqnarray*}
		\arrayin{
			& & (\mu\circ(\nu\cup\xi))(x,y)
			\just={definitions of $\circ$ and $\cup$}
			\sum_{z\in X}\mu(x,z);(\nu(z,y)+\xi(z,y))
			\just={definition of $\sum$ and $z_i\in X, 1\leq i\leq n$}
			\mu(x,z_1);(\nu(z_1,y)+\xi(z_1,y))+\cdots+\mu(x,z_n);(\nu(z_n,y)+\xi(z_n,y))
			\just={(\ref{eqn7})}
			\begin{aligned}
			\mu(x,z_1);\nu(z_1,y)+\mu(x,z_1);\xi(z_1,y)+\dots\\
			+\mu(x,z_n);\nu(z_n,y)+\mu(x,z_n);\xi(z_n,y)
			\end{aligned}
			\just={(\ref{eqn2})
			}
			\begin{aligned}
				\mu(x,z_1);\nu(z_1,y)+\cdots+\mu(x,z_n);\nu(z_n,y)\\
				+\mu(x,z_1);\xi(z_1,y)+\cdots+\mu(x,z_n);\xi(z_n,y)
			\end{aligned}
			\just={definition of $\sum$}
			\sum_{z\in X}\mu(x,z);\nu(z,y)+\sum_{z\in X}\mu(x,z);\nu(z,y)
			\just={definitions of $\circ$ and $\cup$ on fuzzy relations}
			((\mu\circ\nu)\cup(\mu\circ\xi))(x,y)
			}
	\end{eqnarray*}

\noindent
\underline{Axioms (\ref{eqn8}) to (\ref{eqn13})}:
The proof of Axiom (\ref{eqn8}) is analogous. Axiom (\ref{eqn9}) follows straightforwardly, since $\varnothing(x,y)=0$ is the absorbent element of $;$ over $K$.				
Axioms (\ref{eqn10})-(\ref{eqn13}) are proved as in Theorem \ref{theoFSET}, but, of course, taking the definition of composition of fuzzy relations, i.e. $$(\mu\circ\nu)(x,y)=\sum_{z\in X}\mu(x,z);\nu(z,y)$$ for all $\mu$, $\nu$ $\in K^{X\times X}$. As in Theorem \ref{theoFSET}, we only verify axioms (\ref{eqn10}) and (\ref{eqn12}). The validity of 
(\ref{eqn13}) is left for the reader, since the arguments used are essentially the same of (\ref{eqn12}).
~\\

\noindent
\underline{Axiom (\ref{eqn10})}:
\begin{eqnarray*}
	\arrayin{
		& & (\Delta\cup(\mu\circ\mu^*))(x,y)
		\just={definition of $\cup$, $\circ$ and $^*$}
		\Delta(x,y)+\sum_{z\in X}(\mu(x,z);(\sum_{k\geq0} \mu^k(z,y)))
		\just={definition of $\mu$}
		\Delta(x,y)+\sum_{z\in X}(\mu(x,z);
		\mu^0(z,y)+\mu(x,z);\mu(z,y)+\cdots)
	%
	%
		%
		\just={definition of $\sum$ and $z_i\in X, 1\leq i\leq n$}
		\begin{aligned}
			\Delta(x,y)+\mu(x,z_1);
			\mu^0(z_1,y)+\mu(x,z_1);\mu(z_1,y)+\cdots\\
			+\mu(x,z_n);\mu^0(z_n,y)+\mu(x,z_n);\mu(z_n,y)+\cdots
		\end{aligned}
		\just={definition of $\mu^k$}
		\Delta(x,y)+\mu(x,y)+\mu(x,y)+\cdots+\mu(x,y)+\mu(x,y)+\cdots
		\just={(\ref{eqn2}) and (\ref{eqn3})
		}
		\Delta(x,y)+\mu(x,y)+\mu(x,y)+\cdots
		\just={definition of $\mu^k$}
		\mu^*(x,y)
	}
\end{eqnarray*}

\noindent
\underline{Axiom (\ref{eqn12})}:
We assume the left side of the implication of (\ref{eqn12}) for elements of $K$, i.e. $$(\mu\circ\nu)(x,y)\leq\nu(x,y)\Leftrightarrow
\sum_{z\in X}\mu(x,z);\nu(z,y)\leq\nu(x,y)$$ by definition of $\circ$ on fuzzy relations.

\begin{eqnarray}
& & (\mu^*\circ\nu)(x,y)\nonumber
\just={definitions of $\circ$ and $^*$}\nonumber
\sum_{z\in X}\Big(\big(\sum_{k\geq 0}\mu^k(x,z)\big);\nu(z,y)\Big)\nonumber
\just={definition of $\sum$, (\ref{eq:infdist2}) 
and $z_i\in X, 1\leq i\leq n$}
\begin{aligned}\label{step}{}
\mu^0(x,z_1);\nu(z_1,y)+\mu(x,z_1);\nu(z_1,y)+\cdots\\+\mu^0(x,z_n);\nu(z_n,y)+\mu(x,z_n);\nu(z_n,y)+\cdots
\end{aligned}
\end{eqnarray}

%
	
	\noindent 
	Resorting to (\ref{eqn2}) and the hypothesis, the terms of (\ref{step}) are re-organised as follows.
		For $k=0$
	
	\begin{eqnarray*}
			\mu^0(x,z_1);\nu(z_1,y)+\cdots +\mu^0(x,z_n);\nu(z_n,y)\leq \nu(x,y)
		\end{eqnarray*}
		
		\noindent and for $k=1$
		
		\begin{eqnarray*}
				\mu(x,z_1);\nu(z_1,y)+\cdots +\mu(x,z_n);\nu(z_n,y)\leq \nu(x,y).
			\end{eqnarray*}
			
			\noindent Each term $\mu^k(x,z_i);\nu(z_i,y)$, for $k\geq 2$, for each $z_i$, $1\leq i\leq n$, becomes
			
			\begin{footnotesize}
				\begin{equation*}
					(\mu\circ\cdots\circ\mu)(x,z_i);\nu(z_i,y)=\sum_{w^1\in X}\big(\cdots\sum_{w^k\in X}\big(\mu(x,w^k);\mu(w^k,w^{k-1})\big);\cdots ;\mu(w^1,z_i)\big);\nu(z_i,y)
				\end{equation*}
			\end{footnotesize}

			
			\noindent Using (\ref{eq:infdist2}), (\ref{eqn5}) and the hypothesis, we can simplify the expression and prove (\ref{eqn12}).
			As an example, the term for $k=2$, for each $z_i$, $1\leq i\leq k$ is computed as follows; generalisation to other values for $k$ being straightforward.

\begin{eqnarray*}
	\arrayin{
& & \mu^2(x,z_i);\nu(z_i,y)
\just={definition of $\mu^i$ of Definition \ref{defFREL}}
(\mu\circ\mu)(x,z_1);\nu(z_i,y)
\just={definition of $\circ$}
\big(\sum_{w\in X}\big(\mu(x,w);\mu(w,z_i)\big)\big);\nu(z_i,y)
\just={definition of $\sum$ and $w_i\in X, 1\leq i\leq n$}
\big(\mu(x,w_1);\mu(w_1,z_i)+\cdots+\mu(x,w_n);\mu(w_n,z_i)\big);\nu(z_i,y)
\just={(\ref{eq:infdist2}) and (\ref{eqn5})
}
\mu(x,w_1);\big(\mu(w_1,z_i);\nu(z_i,y)\big)+\cdots+\mu(x,w_n);\big(\mu(w_n,z_i);\nu(z_i,y)\big)
\just\leq{$\mu(x,z);\nu(z,y)\leq \nu(z,y)$ for all $x,y,z\in X$ and monotonicity of $;$ and $+$}
\mu(x,w_1);\nu(w_1,y)+\cdots+\mu(x,w_n);\nu(w_n,y)
\just={hypothesis}
\sum_{w\in X}\mu(x,w);\nu(w,y)\leq \nu(x,y)
}
\end{eqnarray*}

\noindent
So, we prove that (\ref{step}) becomes $\nu(x,y)+\cdots+\nu(x,y)$, reduced by (\ref{eqn3}) to $\nu(x,y)$.
~\\
	
\noindent
\underline{Axiom (\ref{eqn14}) (``$\Rightarrow$'')}:
 Let $\sigma$, $\eta$, $\theta$ $\in T^{X\times X}$ and assume
	\begin{eqnarray*}
		\arrayin{
			& & (\sigma\circ\eta)(x,y)\leq\theta(x,y)
			\just \Leftrightarrow {definition of $\circ$}
			\sum_{z\in X}\sigma(x,z);\eta(z,y)\leq\theta(x,y)
			\just \Leftrightarrow {definition of $\sum$ and $z_i\in X, 1\leq i\leq n$}
			\sigma(x,z_1);\eta(z_1,y)+\cdots+\sigma(x,z_n);\eta(z_n,y)\leq\theta(x,y)
		}
	\end{eqnarray*}
	
	Since $\sigma(x,z_i), \eta(z_i,y)\in T^{X\times X}$, there is, at most, one $1\leq i\leq n$ such that $x=z_i$ and $z_i=y$.
	
	So, $\sigma(x,z_1);\eta(z_1,y)+\cdots+\sigma(x,z_n);\eta(z_n,y) = \sigma(x,z_i);\eta(z_i,y)\leq\theta(x,y)$, for the only $1\leq i\leq n$ such that $x=z_i$ and $z_i=y$. Since $\sigma(x,z_i)$, $\eta(z_i,y)$ and $\theta(x,y)$ $\in T$, by (\ref{eqn14}) on $T$, $\sigma(x,z_i);\eta(z_i,y)\leq\theta(x,y)$ implies $\eta(x,y)\leq \sigma(x,y)\to\theta(x,y)$.
	The proof of (``$\Leftarrow$'') is analogous.
	
%
	~\\
	\noindent
	\underline{Axiom (\ref{eqn17})}: 
	The proof of (\ref{eqn17}) is trivial, since $\sigma(x,y)\leq 1=\Delta(x,y)$, for all $\sigma(x,y)\in T^{X\times X}$.
	~\\
	
	\noindent
	\underline{Axiom (\ref{eqn18})}: 
	First observe that
	
	\begin{eqnarray*}
		\arrayin{
	& &	(\sigma\circ\eta)(x,y)
	\just={definition of $\circ$}
	\sum_{z\in X}\sigma(x,z);\eta(z,y)
	\just={definition of $\sum$ and $z_i\in X, 1\leq i\leq n$}
	\begin{aligned}
	\sigma(x,z_1);\eta(z_1,y)+\cdots+\sigma(x,z_n);\eta(z_n,y),\\
	\text{for all } \sigma(x,z_i), \eta(z_i,y)\neq 0 \text{, with } 1\leq i\leq n
\end{aligned}
}
\end{eqnarray*}

Clearly $x=z_i=y$, using the definition of $\sigma(x,y)$, for all $\sigma\in T^{X\times X}$.
Thus, the proof follows directly from (\ref{eqn18}) for elements of $T$, as shown below.
	
	\begin{eqnarray*}
		\arrayin
		{
				& &\eta(x,z_1);\sigma(z_1,y)+\dots+\eta(x,z_n);\sigma(z_n,y)
				\just={definition of $\sum$}
				\sum_{z\in X}\eta(x,z);\sigma(z,y)
				\just={definition of $\circ$}
				(\eta\circ\sigma)(x,y)
			}
	\end{eqnarray*}
	
	

To prove that $\mathbf{FREL(T)}$ is also a I-GKAT, for any complete residuated lattice $\T$, we need to prove axiom (\ref{eq25}).
~\\

\noindent
\underline{Axiom (\ref{eq25})}:
\begin{eqnarray*}
	\arrayin{
		& & (\sigma\circ\sigma)(x,y)
		\just={definition of $\circ$}
		\sum_{z\in X}\sigma(x,y);\sigma(z,y)
		\just={definition of $\sum$ and $z_i\in X, 1\leq i\leq n$}
		\sigma(x,z_1);\sigma(z_1,y)+\cdots+\sigma(x,z_n);\sigma(z_n,y)
		}
\end{eqnarray*}

Again, $\sigma(x,y)+\sigma(x,z_1);\eta(z_1,y)+\cdots+\sigma(x,z_n);\eta(z_n,y)$ reduces to $\sigma(x,z_i);\sigma(z_i,y)$, for the only $1\leq i\leq n$ such that $x=z_i=y$.
But $\sigma(x,z_i);\sigma(z_i,y)=\sigma(x,y)$, by (\ref{eq25}), since $\sigma(x,z_i), \sigma(z_i,y)\in T$.
~\\

\section{Proof of Theorem 6}\label{sec:th6}

	It was already proved by Kozen \cite{Kozen1994} that the structure	
	\[(M(n,K),+,;,^*,0_n,I_n)\]
	forms a Kleene algebra. Then, it remains to prove that	
	\[(\Delta(n,T),+,;\rightarrow,0_n,I_n)\]	
	is the subalgebra of Definition 2, i. e. satisfies the axioms (\ref{eqn14})-(\ref{eqn18}).
	\\
	\\
	Let $A$=
	$	\begin{bmatrix}
	a_{11} & 0 & \cdots & 0\\
	0 & a_{22} & \cdots & 0\\
	\hdotsfor{4}\\
	0 & 0 & \cdots & a_{nn}
	\end{bmatrix}$
	, $B$=
	$\begin{bmatrix}
	b_{11} & 0 & \cdots & 0\\
	0 & b_{22} & \cdots & 0\\
	\hdotsfor{4}\\
	0 & 0 & \cdots & b_{nn}
	\end{bmatrix}$\\
	and $C$=
	$\begin{bmatrix}
	c_{11} & 0 & \cdots & 0\\
	0 & c_{22} & \cdots & 0\\
	\hdotsfor{4}\\
	0 & 0 & \cdots & c_{nn}
	\end{bmatrix}
	$\\ be elements of $\Delta(n,T)$.
	
	For (\ref{eqn14}) we prove that $A;B+C=C\Rightarrow B+A\rightarrow C=A\rightarrow C$.
 Using the definitions of the operators, we obtain

	\begin{align*}
	& \; \; \; \begin{bmatrix}
	a_{11} & 0 & \cdots & 0\\
	0 & a_{22} & \cdots & 0\\
	\hdotsfor{4}\\
	0 & 0 & \cdots & a_{nn}
	\end{bmatrix}
	;
	\begin{bmatrix}
	b_{11} & 0 & \cdots & 0\\
	0 & b_{22} & \cdots & 0\\
	\hdotsfor{4}\\
	0 & 0 & \cdots & b_{nn}
	\end{bmatrix}
	+
	\begin{bmatrix}
	c_{11} & 0 & \cdots & 0\\
	0 & c_{22} & \cdots & 0\\
	\hdotsfor{4}\\
	0 & 0 & \cdots & c_{nn}
	\end{bmatrix}
	\\
	= & \; \; \; 	
	\begin{bmatrix}
	c_{11} & 0 & \cdots & 0\\
	0 & c_{22} & \cdots & 0\\
	\hdotsfor{4}\\
	0 & 0 & \cdots & c_{nn}
	\end{bmatrix}
	\end{align*}
	
	\noindent
	which is equivalent to 

$$
	\begin{bmatrix}
	a_{11};b_{11}+c_{11} & 0 & \cdots & 0\\
	0 & a_{22};b_{22}+c_{22} & \cdots & 0\\
	\hdotsfor{4}\\
	0 & 0 & \cdots & a_{nn};b_{22}+c_{nn}
	\end{bmatrix}
	=
	\begin{bmatrix}
	c_{11} & 0 & \cdots & 0\\
	0 & c_{22} & \cdots & 0\\
	\hdotsfor{4}\\
	0 & 0 & \cdots & c_{nn}
	\end{bmatrix}
$$

	\noindent
	In order for two matrices to be equal, their elements must be equal in the corresponding positions. So, the assumption is
	
	$$
	\begin{cases}
	a_{11};b_{11}+c_{11}=c_{11}\\
	a_{22};b_{22}+c_{22}=c_{22}\\
	\cdots\\
	a_{nn};b_{nn}+c_{nn}=c_{nn}
	\end{cases}
	$$
	
	\noindent	We have to prove that
	
		\begin{align*}
	& \; \; \; \begin{bmatrix}
	b_{11} & 0 & \cdots & 0\\
	0 & b_{22} & \cdots & 0\\
	\hdotsfor{4}\\
	0 & 0 & \cdots & b_{nn}
	\end{bmatrix}
	+
	\begin{bmatrix}
	a_{11} & 0 & \cdots & 0\\
	0 & a_{22} & \cdots & 0\\
	\hdotsfor{4}\\
	0 & 0 & \cdots & a_{nn}
	\end{bmatrix}
	\rightarrow
	\begin{bmatrix}
	c_{11} & 0 & \cdots & 0\\
	0 & c_{22} & \cdots & 0\\
	\hdotsfor{4}\\
	0 & 0 & \cdots & c_{nn}
	\end{bmatrix}\\
	= & \; \; \; 	
	\begin{bmatrix}
	a_{11} & 0 & \cdots & 0\\
	0 & a_{22} & \cdots & 0\\
	\hdotsfor{4}\\
	0 & 0 & \cdots & a_{nn}
	\end{bmatrix}
	\rightarrow
	\begin{bmatrix}
	c_{11} & 0 & \cdots & 0\\
	0 & c_{22} & \cdots & 0\\
	\hdotsfor{4}\\
	0 & 0 & \cdots & c_{nn}
	\end{bmatrix}
	\\
	\Leftrightarrow & \; \; \;
	\begin{bmatrix}
	b_{11}+a_{11}\rightarrow c_{11} & 0 & \cdots & 0\\
	0 & b_{22}+a_{22}\rightarrow c_{22} & \cdots & 0\\
	\hdotsfor{4}\\
	0 & 0 & \cdots & b_{22}+a_{22}\rightarrow c_{22}
	\end{bmatrix}\\
	= & \; \; \; 
	\begin{bmatrix}
	a_{11}\rightarrow c_{11} & 0 & \cdots & 0\\
	0 & a_{22}\rightarrow c_{22} & \cdots & 0\\
	\hdotsfor{4}\\
	0 & 0 & \cdots & a_{nn}\rightarrow c_{nn}
	\end{bmatrix}	
	\end{align*}
	
	\noindent
	i. e.
	
	$$\begin{cases}
	b_{11}+a_{11}\rightarrow c_{11}=a_{11}\rightarrow c_{11}\\
	b_{22}+a_{22}\rightarrow c_{22}=a_{22}\rightarrow c_{22}\\
	\cdots\\
	b_{nn}+a_{nn}\rightarrow c_{nn}=a_{nn}\rightarrow c_{nn}
	\end{cases}$$	
	
	\noindent	 
	But, since $a_{ij}, b_{ij}, c_{ij}\in T$ for all $1\leq i,j\leq n $ it is verified by axiom (\ref{eqn14}) of GKAT that
	
	$$\begin{cases}
	a_{11};b_{11}+c_{11}=c_{11}\Rightarrow b_{11}+a_{11}\rightarrow c_{11}=a_{11}\rightarrow c_{11}\\	 
	a_{22};b_{22}+c_{22}=c_{22}\Rightarrow
	b_{22}+a_{22}\rightarrow c_{22}=a_{22}\rightarrow c_{22}\\
	\cdots\\
	a_{nn};b_{nn}+c_{nn}=c_{nn}\Rightarrow
	b_{nn}+a_{nn}\rightarrow c_{nn}=a_{nn}\rightarrow c_{nn}
	\end{cases}$$
	
	The proof for $\Leftarrow$ is similar.
	The proofs of axioms (\ref{eqn17}) and (\ref{eqn18}) are analogous, using the definitions of the operators over elemtents of $\Delta(n,T)$.
	Note, in particular, the proof of axiom (\ref{eqn18}).
	It is well known that the multiplication of matrices is not commutative. However, since axiom (\ref{eqn18}) is only applied to elements of $\Delta(n,T)$, that is, diagonal matrices, and the multiplication of diagonal matrices is commutative, this axiom is valid for all GKAT.
	
	To prove that this also forms a I-GKAT, it suffices to show the validity of (\ref{eq25}). 
	The proof is similar to the one presented for GKAT, for all $A, B, C\in \Delta(n,T)$, using the definitions of the operators over elements of $\Delta(n,T)$.

\section{Proof of Theorem 7}\label{sec:th7}

	To prove the equivalence, we need the following identities:

\begin{eqnarray}
	p;(q;p)^* &=&(p;q)^*;p\label{eq31}\\
	p^*;(q;p^*)^*&=&(p+q)^*\label{eq32}
\end{eqnarray}

\noindent which are derivable from the axioms of Kleene algebra and were proved in \cite{Kozen1994}.


%

\noindent
In order to prove the equivalence (\ref{eq29})$\Leftrightarrow$(\ref{eq30}), let us start by translating both programs to the language of Kleene algebra. Program (\ref{eq29}) becomes

\begin{equation}
	(b;p;(c;q)^*;(c\to 0))^*;(b\to 0),\label{eq33}
\end{equation}

\noindent and (\ref{eq30}) becomes\footnote{As in Kozen's paper \cite{Kozen1997}, we 
interpret the program $\textbf{if} \; b \;\textbf{then} \;p$ as an abbreviation for a conditional test with the dummy \textbf{else} clause  i.e., as the program $b;p + \bar{b}$ ($b;p + b\rightarrow 0$ in our setting).}
\begin{equation}
	b;p;((b+c);(c;q+(c\to 0);p))^*;((b+c)\to 0)+(b\to 0)\label{eq34} 
\end{equation}

\noindent Simplifying (\ref{eq33}),

\begin{eqnarray*}
	\arrayin{
		& &(b;p;(c;q)^*;(c\to 0))^*;(b\to 0)
		\just={(\ref{eqn10})}
		(1+b;p;(c;q)^*;(c\to 0);(b;p;(c;q)^*;(c\to 0))^*);(b\to 0)
		\just={(\ref{eqn8})}
		(b\to 0)+b;p;(c;q)^*;(c\to 0);(b;p;(c;q)^*;(c\to 0))^*;(b\to 0)
		\just={(\ref{eq31})}
		(b\to 0)+b;p;(c;q)^*;(c\to 0);(b;p;(c;q)^*)^*;(c\to 0);(b\to 0)
		}
\end{eqnarray*}

\noindent For (\ref{eq34}), the sub expression $(b+c);(c;q+(c\to 0);p)$ becomes

\begin{eqnarray*}
	\arrayin{
		& & (b+c);(c;q+(c\to 0);p)
		\just={(\ref{eqn7})}
		b;c;q+b;(c\to 0);p+c;c;q+c;(c\to 0);p
		\just={(\ref{eq25}), $a;(a\to 0)=0$, (\ref{eqn9}) and (\ref{eqn4})}
		b;c;q+b;(c\to 0);p+c;q
		\just={(\ref{eqn3}) and (\ref{eqn18})}
		b;c;q+c;q+(c\to 0);b;p
		\just={(\ref{eqn8})}
		(b+1);c;q+(c\to 0);b;p
		}
\end{eqnarray*}

\noindent Moreover, $(b+c)\to 0=(b\to 0);(c\to 0)$, by (\ref{eq:demorgan}). Applying these transformations on (\ref{eq34}), we obtain

\begin{equation*}
	b;p;(c;q+(c\to 0);b;p)^*;(b\to 0);(c\to 0)+(b\to 0)
\end{equation*}

\noindent Now, we need to prove that

\begin{align*}
	(b\to 0)+b;p;(c;q)^*;(c\to 0);(b;p;(c;q)^*)^*;(c\to 0);(b\to 0)=\\
	=b;p;(c;q+(c\to 0);b;p)^*;(b\to 0);(c\to 0)+(b\to 0)
\end{align*}

\noindent But, by monotonicity of operators $+$ and $;$, this expression is equivalent to

\begin{align*}
	(c;q)^*;(c\to 0);(b;p;(c;q)^*)^*=(c;q+(c\to 0);b;p)^*
\end{align*}

\noindent which is just an instance of the denesting rule (\ref{eq32}).
